# Correlated chemical and displacive disorder in a lithium-rich disordered rocksalt revealed by electron diffuse scattering

Sam Y. Thompson[a], Kristoffer A. H. Støckler[a], Nikolaos Zacharakis[a], Rebekka Klemmt[a], Ella M. Schmidt[b], Sandro Wieser[c], Nico Unglert[c], Jørgen Skibsted[a], Espen D. Bøjesen[a], Georg K. H. Madsen[c], Bo B. Iversen[a,*], Dorthe B. Ravnsbæk[a,*]

[a]Center for Sustainable Energy Materials, Department of Chemistry, Aarhus University, Langelandsgade 140, Aarhus C, DK-8000, Denmark
[b]Faculty of Geosciences, MARUM and MAPEX, University of Bremen, Klagenfurter Straße 2-4, DE-28359 Bremen, Germany
[c]Institute of Materials Chemistry, TU Wien, 1060 Vienna, Austria

*Corresponding author e-mails: bo@chem.au.dk, dorthe@chem.au.dk

## Abstract

Lithium transport in disordered rocksalt cathodes depends on local cation arrangements and atomic distortions that are hidden from average-structure analysis. Resolving both contributions experimentally remains challenging, particularly in powders where directional information is lost. Here, we use three-dimensional electron diffuse scattering and three-dimensional difference pair distribution function (3D-ΔPDF) analysis to resolve occupational and displacive correlations in the model lithium-rich rocksalt $Li_{2.5}TiO_3F_{0.5}$. We identify an unusual combination of like-cation first-neighbour and unlike-cation second-neighbour preferences, together with correlated anion displacements towards Ti and accompanying cation displacements. A minimal pair-interaction model incorporating these displacements reproduces the principal diffuse-scattering and 3D-ΔPDF features and improves agreement with independent powder-PDF measurements. The experimentally constrained cation configurations exhibit an above-random population of 0-TM lithium-migration environments. These results establish electron 3D-ΔPDF as a route to constrain atomistic models incorporating both chemical order and local relaxation, providing a foundation for assessing their combined influence on lithium migration in disordered rocksalt materials.

Cation-disordered rocksalt (DRX) cathodes are promising candidates for next-generation lithium-ion batteries, combining high lithium-storage capacities with broad compositional flexibility. Occupational disorder in the rocksalt structure permits the incorporation of a wide range of redox-active transition metals,[1,2] providing compositional flexibility that can reduce reliance on critical elements such as cobalt and nickel. Lithium transport in DRX cathodes is governed by local cation configurations that are hidden by their apparently simple average structure.[3] Lithium and transition metal ions occupy the octahedral sites of a cubic close-packed anion framework, comprising oxygen and, in oxyfluorides, fluorine. Lithium migration occurs between adjacent octahedral sites through intermediate tetrahedral sites, with activation barriers strongly dependent on the surrounding cation configuration. Tetrahedra that face-share with four Li-occupied octahedra, termed 0-TM environments, generally provide the lowest migration barriers. As the number of high-valent transition metals surrounding the tetrahedral site increases, their electrostatic repulsion with the migrating lithium ion increases, raising the activation barrier giving the hierarchy 0-TM < 1-TM < 2-TM.[2] Therefore, one design method for new DRX materials is to engineer structures with increased 0-TM population and connectivity.

The distribution of ions within the rock salt framework is generally not random.[3–5] Local charge compensation, bonding preferences, ionic-size mismatch and electronic effects such as Jahn–Teller distortions can generate short-range order, often favouring local Li–TM mixing.[6,7] Such correlations can reduce the population or connectivity of 0-TM environments relative to random cation mixing, motivating strategies to suppress ordering, including the use of high-entropy cation chemistries.[7,8] Short-range order is not, however, intrinsically detrimental. Within an equiatomic pseudobinary fcc model, Liu et al. identified interaction regimes in which a preference for like-cation first neighbours increases the 0-TM population, while competing second-neighbour interactions oppose macroscopic segregation.[9] Ouyang et al. showed that coupled cation–anion ordering driven by Li–F bonding preferences can likewise increase the 0-TM population above the random limit.[10,11] These studies highlight the importance of the specific correlations present, rather than simply the extent of departure from random mixing. Atomic displacements provide a further contribution to short-range order and may lower migration barriers and enable transport through some 1-TM environments extending the accessible network beyond that predicted by considering 0-TM channels alone.[12,13] Displacements also account for much of the deviation from the average structure observed in powder pair distribution function (PDF) analyses.[4] Understanding how local structure affects lithium transport therefore requires detailed descriptions of short-range order including both occupational correlations and atomic displacements.

The deviation from the average (random) structure generates diffuse scattering between Bragg reflections. Although short-range order in DRX materials has been studied extensively using cluster-expansion methods,[3,7,10,14,15] experimental comparisons generally remain limited to qualitative agreement between predicted and measured diffuse-scattering morphologies.[3,5,8,16–19] Common patterns have been interpreted in terms of local tetrahedral, octahedral or cubic environments whose compositions are constrained towards the bulk composition (orange, purple and red polyhedra, respectively, in Fig.1b).[20,21,5] Such descriptions provide useful classifications, but do not by themselves resolve the occupational and displacive correlations underlying the measured intensity.

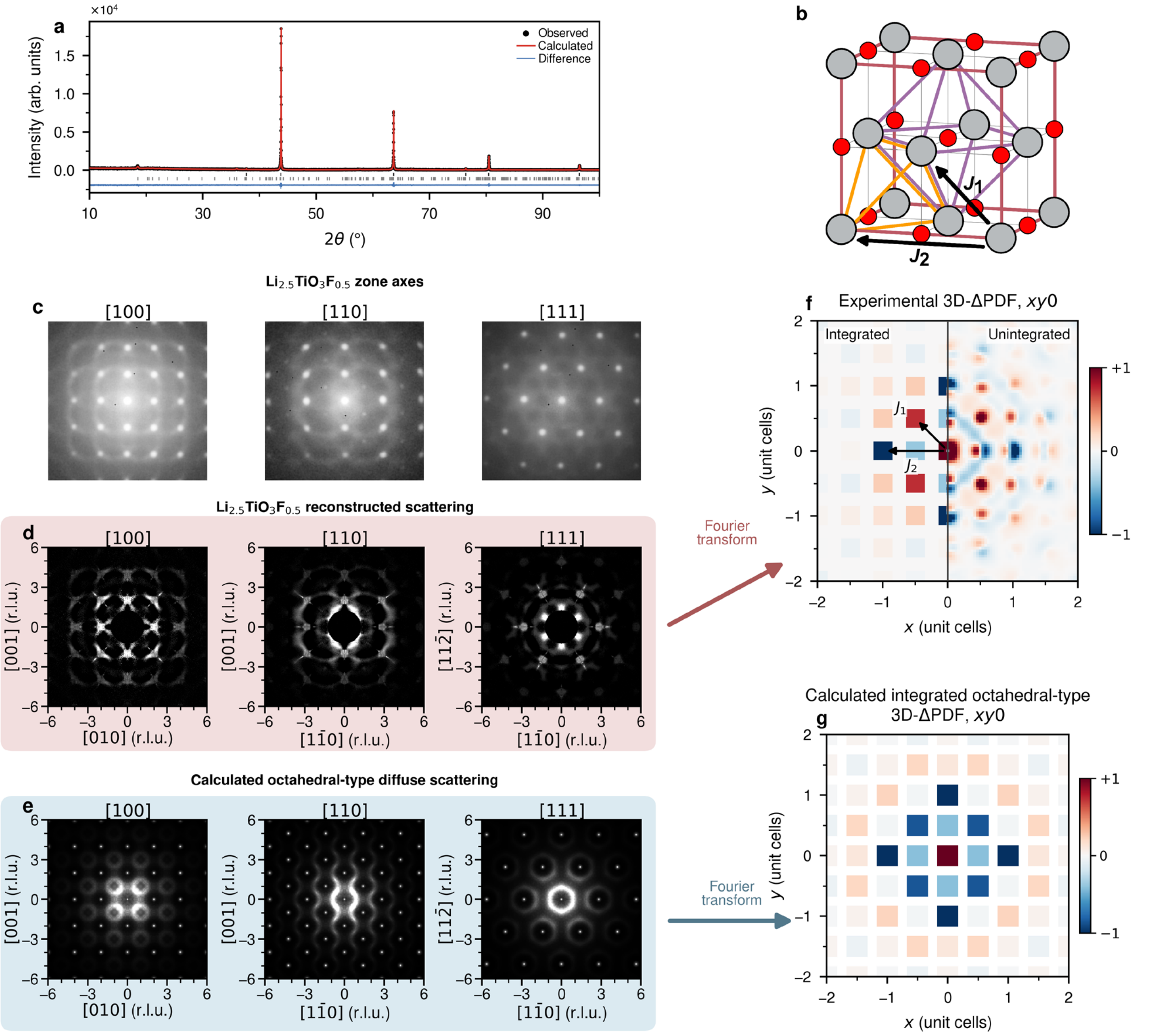

**Fig. 1 | Experimental observations of $Li_{2.5}TiO_3F_{0.5}$.** a, Powder X-ray diffraction data and Rietveld refinement of $Li_{2.5}TiO_3F_{0.5}$. The peak at ~18.5° is due to a small $Li_2TiO_3$ impurity. b. Rocksalt crystal structure showing the tetrahedral, octahedral and cubic local polyhedra in orange, purple and red, respectively and which nearest neighbour sites the interactions $J_1$ and $J_2$ operate on. In LTOF, lithium and titanium are disordered over the grey sites and oxygen and fluorine are disordered over the red sites. c, Electron-diffraction patterns recorded for LTOF along the [100], [110] and [111] zone axes. d, Sections through the three-dimensional diffuse-scattering volume reconstructed from static frame rotation electron diffraction, normal to [100], [110] and [111]. e, For comparison, calculated electron diffraction zone axes for an example model of octahedral-type ordering adopted by common DRX compositions such as $Li_{1.2}Mn_{0.4}Ti_{0.4}O_2$ and $Li_{1.3}Mn_{0.4}Nb_{0.3}O_2$. f, Experimental 3D-ΔPDF section at z = 0 for $Li_{2.5}TiO_3F_{0.5}$ (right) and the experimental 3D-ΔPDF after integrating the intensity in 7x7x7 voxels around each ideal interatomic vector (atomic position) (left). Red and blue denote positive and negative 3D-ΔPDF intensity, respectively, representing deviations from the average-structure pair correlations. Real-space coordinates are expressed in conventional cubic unit cells (1 r.l.u. ~ 4.1 Å). g, For comparison, integrated 3D-ΔPDF of the model in e.

Hence, there is a wealth of underutilized information contained within the diffuse scattering of DRX materials. To address this, we apply three dimensional difference pair distribution function analysis (3D-ΔPDF)[22] to a DRX material. Unlike conventional one-dimensional PDF analysis, which provides a radial distribution of interatomic distances but loses directional information, 3D-ΔPDF retains the full three-dimensional character of pair correlations. By Fourier transforming the diffuse-scattering intensity after removal of the Bragg reflections, the method isolates correlations associated with deviations from the average crystal structure, providing direct directional insight into short-range order and correlated atomic displacements. Originally developed for single-crystal X-ray and neutron scattering, 3D-ΔPDF has so far been applied to battery materials in only a few studies.[23–25] A promising recent development is its extension to electron diffraction (ED).[26,27] By combining 3D-ΔPDF analysis with ED, we can investigate individual crystallites directly from functional DRX powders, allowing short-range order and correlated atomic displacements to be studied at application-relevant crystallite sizes without requiring the growth of larger crystals that may not preserve the same local structure.

Here, we investigate $Li_{2.5}TiO_3F_{0.5}$ (LTOF) as a model fluorinated DRX material using static-frame rotation electron diffraction and the 3D-ΔPDF analysis. Its binary Li/Ti cation sublattice limits the number of distinct occupational correlations, while the strong electron-scattering contrast between Li and Ti facilitates their identification. The nominally $d^0$ $Ti^{4+}$ chemistry also simplifies the interpretation of local distortions by avoiding mixed transition-metal valence and conventional Jahn–Teller-active cations such as $Mn^{3+}$.[28–30]. LTOF thus provides a comparatively simple system for examining occupational and displacive correlations relevant to the broader DRX family.

We find that LTOF exhibits a markedly different diffuse scattering morphology compared to previously reported DRX materials. The 3D-ΔPDF resolves the underlying occupational correlations, revealing a preference for like-cation first neighbours and unlike-cation second neighbours. It also identifies directional anion displacements towards Ti and accompanying correlated cation displacements, providing a spatially resolved description of local relaxation. A minimal pair-interaction model incorporating these displacement modes reproduces the principal experimental features, while independent powder-PDF measurements demonstrate the importance of including displacements in the structural description. The resulting cation arrangement increases the 0-TM population above the random limit, providing direct experimental evidence that short-range order can favour, rather than necessarily suppress, locally favourable lithium-migration environments. Broadly, the results establish an experimental basis for modelling both the chemical arrangements that define migration networks and the local distortions that may modify migration barriers in DRX materials.

## Results

**Synthesis and characterization.** LTOF was synthesized by solid state synthesis. Rietveld refinement of the powder X-ray diffraction data for the main DRX phase, using a cubic unit cell with space group $Fm\bar{3}m$, yielded a unit cell parameter of $a$ = 4.131251(11) Å. Small impurity peaks were fit using a $C2/c$ layered $\beta$-$Li_2TiO_3$ phase refining to a content of ~7 wt% (Fig. 1a). Seven individual crystallites dispersed from the powder sample were first examined by electron diffraction without static frame rotation. Fig. 1c shows representative patterns indexed along the [100], [110] and [111] zone axes. In the [100] pattern, overlapping rings of diffuse intensity are centred on the ($h$,$k$ = odd), ($l$ = 0) reciprocal-lattice positions. Similar overlapping rings are observed in the equivalent [110] section, whereas the [111] pattern contains star-shaped diffuse features extending from the Bragg reflections. The ring- and star-shaped diffuse features differ markedly from the diffuse signatures often reported in the literature for disordered rocksalt materials; Fig. 1e and g shows a simulation of the typical diffuse scattering observed and expected 3D-ΔPDF, respectively, for octahedral-type short-range order as observed in compositions such as $Li_{1.2}Mn_{0.4}Ti_{0.4}O_2$ and $Li_{1.3}Mn_{0.4}Nb_{0.3}O_2$.[3,16,17]

### 3D-ΔPDF resolves occupational and displacive correlations

Static frame rotation electron diffraction datasets were collected from isolated crystallites of $Li_{2.5}TiO_3F_{0.5}$ and reconstructed into a three-dimensional reciprocal-space volume. [100], [110] and [111] slices through this volume are shown in Fig. 1d. Small crystallites were selected to reduce dynamical-scattering effects, while spatial isolation from neighbouring particles and the carbon-grid edge minimized overlapping diffraction and background contributions. Two data collections were performed on two separate crystallites and the dataset whose rotation passed through fewer zone axes was used for the following analysis. After removal of the Bragg intensities, the remaining diffuse-scattering volume was Fourier transformed to obtain the 3D-ΔPDF. Supplementary Fig. 10 shows the reproducibility between the two datasets for the diffuse scattering reconstruction and the resulting 3D-ΔPDFs.

In synchrotron X-ray 3D-ΔPDF analysis, Bragg reflections are typically removed using small spherical masks. In electron diffraction, inelastic scattering can contribute intensity around Bragg reflections, which can be reduced by post-column energy filtering.[31,32] In absence of an energy filter, our electron diffraction data contains streaks of intensity extended from the Bragg reflections well beyond the regions that would be covered by small spherical masks. Without careful removal of these Bragg features, the resulting 3D-ΔPDF will be affected by contributions from the average structure. Here, we use a seed-skewness algorithm to determine the extent of the Bragg peaks.[33,34] This method is detailed in Supplementary Methods 2.

The 3D-ΔPDF contains contributions from both occupational and displacive correlations. Due to the higher scattering power of $Ti^{4+}$ compared to $Li^{+}$ and the near-identical scattering power of $O^{2-}$ and $F^{-}$, the contrast creating the diffuse scattering, which is observable in the 3D-ΔPDF, is that between $Ti^{4+}$ and the anions (which we will label X from here on) and $Ti^{4+}$ and $Li^{+}$. To emphasize the occupational contribution, the 3D-ΔPDF intensity was integrated within equivalent volumes surrounding each ideal interatomic-vector position. Dipolar features arising from small displacements, including size effects, approximately cancel under this integration, whereas the net intensity associated with occupational correlations is retained.[35] The integrated 3D-ΔPDF (Fig. 1f) shows positive intensity at the first-neighbour cation vector (½, ½, 0), indicating an excess of like-cation pairs relative to the average structure. This corresponds to a preference of first neighbour Ti–Ti and Li–Li pairs. By contrast, the negative intensity at the second-neighbour cation vector (1,0,0) indicates an excess of unlike Ti–Li pairs. Due to the near identical scattering power of O/F on the anion lattice, the (½, 0, 0) feature should integrate to zero and the slight negative feature that we do observe at this position is an estimate of the integral error. Hence, we can be confident in the signals we observe at (½, ½, 0) and (1, 0, 0) as they are much stronger than the (½, 0, 0) signal.

The unintegrated 3D-ΔPDF sections reveal substantial additional displacive correlations. At the nearest cation–anion vector (½, 0, 0), the asymmetric positive and negative lobes are consistent with displacement of the local position of X towards Ti and away from Li. This displacement direction is consistent with the smaller effective ionic radius of octahedrally coordinated $Ti^{4+}$ relative to $Li^{+}$, and hence with shorter Ti-X than Li-X distances.[36] The first-neighbour cation vector (½, ½, 0) contains a strong occupational feature but comparatively little displacive character, whereas the second-neighbour vector (1,0,0) contains both occupational and displacive contributions.

Together, the 3D-ΔPDF reveals an unusual combination of like-cation first-neighbour correlations, unlike-cation second-neighbour correlations as well as correlated anion displacements towards Ti. These occupational and displacive correlations account for the diffuse-scattering morphology.

### 3D-ΔPDF-informed structural model

To quantify the diffuse scattering shapes observed for LTOF and DRX materials in general, we used a minimal pair-interaction model like that used by Roth et al.[37], and Schmidt and Simonov.[38] An extensive study of interaction space in fcc materials, including a four-body $Li_4$ interaction, has recently been performed, however, considering pair interactions for the first and second nearest neighbours is sufficient for our purposes as they were found to account for the experimentally observed pair correlations.[9] First- and second-neighbour interactions $J_1$ and $J_2$, were defined such that positive values favour unlike Li-Ti pairs and contribute to the energy of the system by:

$$\mathcal{H} = J_1 \sum_{i,j}^{12NN} n_i n_j + J_2 \sum_{i,j}^{6NNN} n_i n_j \quad (1)$$

where each cation site $i$, has an occupation variable $n_i$, which was defined as 1 for Ti and 0 for Li. NN and NNN denote first and second nearest neighbours respectively.

The signs of the experimentally observed cation correlations constrain the interaction model to the $J_1/T < 0$, $J_2/T > 0$ regime. Monte Carlo configurations were generated over a two-dimensional parameter grid spanning $-1 \leq J_1/T \leq 0$ and $0 \leq J_2/T \leq +1$ with increments of 0.25, and the calculated diffuse scattering and 3D-ΔPDFs were compared qualitatively to those from experiment (Supplementary Fig. 7). Several neighbouring parameter combinations around $J_1/T = -0.5$ and $J_2/T = +0.5$ reproduced the principal experimental features. This parameter combination was therefore selected as a representative occupational model for the subsequent displacement modelling. Because comparable agreement was obtained over a broad region of parameter space, the selected values should not be interpreted as a unique determination of the underlying physical pair-interaction energies.

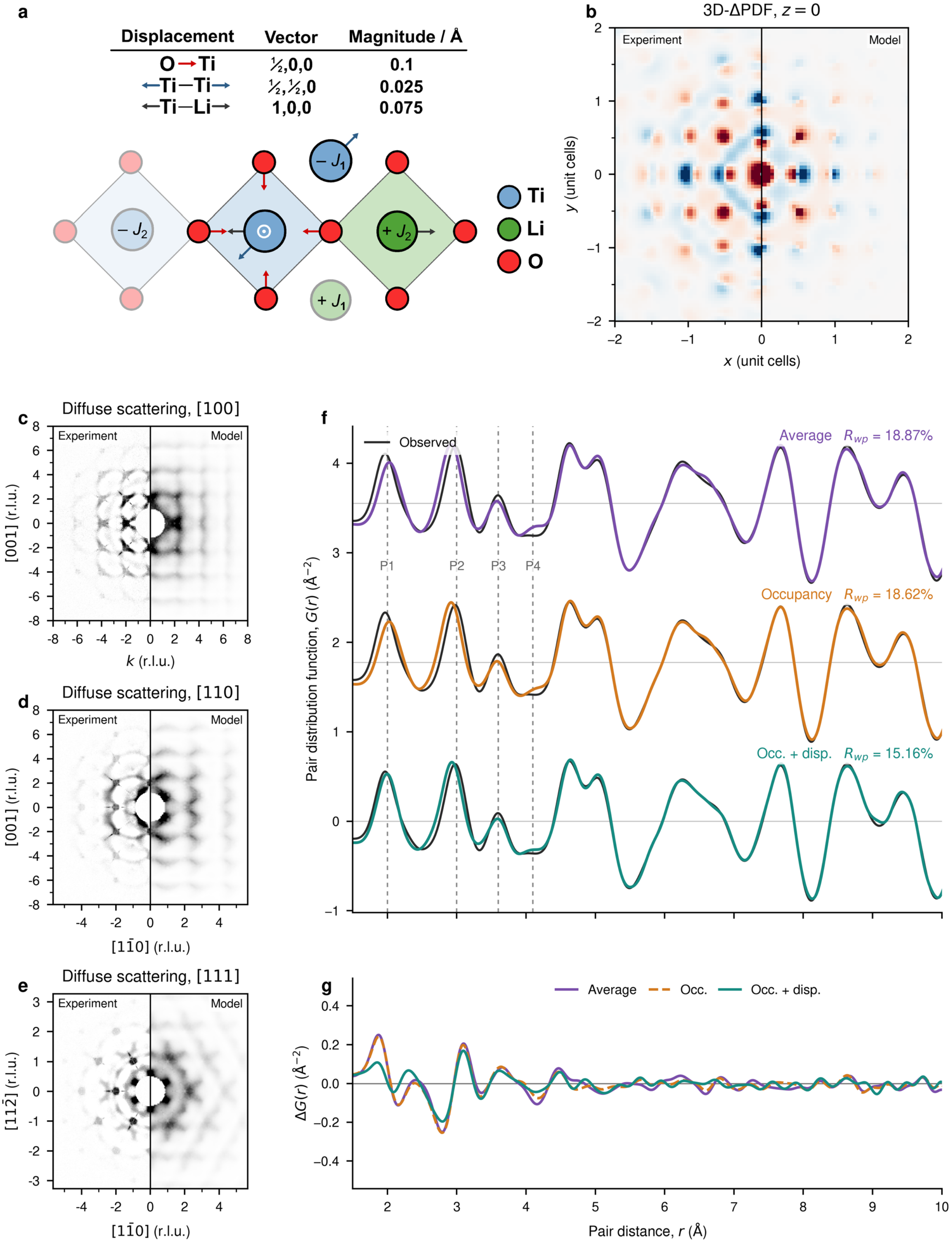

**Fig. 2 | Validation of the 3D-ΔPDF-informed structural model against 3D-ΔPDF and 1D-PDF data.** a, Schematic of the displacive model. b, 3D-ΔPDF section at z = 0 with positive and negative intensity shown in red and blue, respectively. c,d,e, Diffuse-scattering sections normal to [100] (c), [110] (d), [111] (e), with left showing the reconstructed electron diffraction data and right showing the model described in the text. Darker contrast indicates higher intensity. f, Experimental powder PDF and fits obtained using the average (purple), occupation-only large box model (orange) and occupation-plus-displacement large-box model (teal). g, Corresponding difference curves ($G_{obs} - G_{calc}$), shown on a common scale.

Three displacement modes were introduced based on the positions and orientations of the experimental 3D-ΔPDF features: displacement along [100] of X towards neighbouring Ti, separation of first-neighbour Ti-Ti pairs and separation of second-neighbour Li-Ti pairs. The size-effect displacement observed at (½,0,0) is the dominant displacive feature and accounts for much of the change in diffuse scattering relative to the occupational-only model (Supplementary Fig. 8). Varying the displacement magnitude of X towards Ti from 0 to 0.2 Å shows that a displacement of around 0.1 Å reproduces many of the reciprocal-space experimental features (Supplementary Fig. 11). In particular, the crosses centred on {200} and {220} reflections in the [100] zone-axis pattern and the stars centred on {220} reflections in the [111] zone-axis pattern evolve into arcs resembling those observed experimentally. Displacements greater than 0.1 Å produce excessive elongation of the (½,½,0) feature in the 3D-ΔPDF and poorer agreement with the experimental diffuse-scattering intensity distribution. The calculated diffuse-scattering patterns are comparatively insensitive to the magnitudes of the cation displacements, which were therefore selected by comparing the calculated and experimental 3D-ΔPDF features. Each ion in second-neighbour Li–Ti pairs was displaced by 0.075 Å away from its partner, while each ion in first-neighbour Ti–Ti pairs was displaced by 0.025 Å. Representative sweeps of these displacement magnitudes are shown in Supplementary Figs. 12 and 13, respectively. This procedure yields a minimal representative model consistent with the experimental pair correlations, rather than a unique atomistic refinement. Comparison of this model to the experimental diffuse scattering and 3D-ΔPDF are shown in Fig. 2b-e, and show a good match, accounting for most features.

### Model comparison to powder PDF

The structural model has been well-validated against the 3D-ΔPDF and the diffuse scattering, however a further validation can be made against the powder 1D-PDF. First, a Rietveld-like small-box refinement was performed against the average $Fm\bar{3}m$ structure. Here, and in all following PDF refinements, $\beta$-$Li_2TiO_3$ was included as a secondary phase, with the relative phase scales constrained to the PXRD-derived content of 7 wt%. Individual refinements for LTOF and $\beta$-$Li_2TiO_3$ can be seen in Supplementary Fig. 6. Initially, refinement was performed from 1 to 60 Å and showed an excellent fit above ~5 Å. Significant deviations from the average structure are evidenced below 5 Å, as expected, due to the presence of short-range order. This is consistent with the 3D-ΔPDF in Fig. 2 as strong correlations are not observed past ~ 6 Å which also demonstrates that the crystallites selected are representative of the bulk. The refinement was repeated from 1 to 10.5 Å to study the short-range correlations lacking from the average structure description. This achieved an $R_{wp}$ of 18.87% and the LTOF cell parameter refined to 4.1306(9) Å. The resulting fit is shown in Fig. 2f (top).

Fig. 2f compares the experimental PDF with those calculated from the average structure and the two large-box models developed from the 3D diffuse-scattering analysis with the residuals show in Fig. 2g. Four main features poorly fit by the average structure are indicated by the vertical grey dashed lines in Fig. 2f. Peak 1, at approximately 2.0 Å, is attributed to the nearest-neighbour cation–anion correlations. Its experimental maximum is shifted by approximately 0.1 Å towards lower $r$ relative to the average model, consistent with the shorter local Ti–X distances inferred from the 3D-ΔPDF. Peak 2, at approximately 3.0 Å, contains the first-neighbour cation–cation and anion–anion correlations. Its Ti-weighted contribution occurs at slightly higher $r$ than predicted by the average structure, consistent with the correlated separation of first-neighbour Ti–Ti pairs identified in the displacement model. Peak 3, at approximately 3.6 Å, contains the second cation–anion coordination shell and exhibits a very slight elongation relative to the average structure as well as underfit intensity. Peak 4, at approximately 4.1 Å, contains the second-neighbour cation–cation and anion–anion correlations. The average model overestimates its intensity, consistent with non-random pair populations at this distance.

The occupation-only model, generated by the Monte Carlo simulation, produces only a minor improvement relative to the average structure ($R_{wp}$ = 18.62%). Applying the displacement model to the same cation configuration results in a marked reduction in $R_{wp}$ (to 15.16%) and in the magnitude of the low-$r$ residuals. This observation is consistent with the findings of Szymanski et al.[4] that local displacive deviations are the major contributor to low-$r$ residuals in powder PDF data of disordered rocksalt materials. The substantially greater improvement obtained upon incorporating atomic displacements shows that reproducing the cation distribution alone is insufficient to describe the local structure of LTOF. The directional information supplied by electron 3D-ΔPDF therefore provides an essential complement to powder PDF, enabling a structural model that captures both chemical short-range order and the accompanying local relaxation.

## Discussion

### Relating LTOF to previously reported DRX short-range order

The combination of like-cation first-neighbour and unlike-cation second-neighbour correlations in LTOF

differs from the short-range order commonly reported for DRX cathodes e.g. $Li_{1.2}Ti_{0.4}Mn_{0.4}O_2$, $Li_{1.2}Zr_{0.4}Mn_{0.4}O_2$, $Li_{1.3}Nb_{0.3}Mn_{0.4}O_2$. To understand this distinction, we again use the pair-interaction framework to examine the relationship between the experimentally informed LTOF model and the tetrahedral-, octahedral- and cube-like diffuse-scattering morphologies associated with previously reported DRX short-range order. Specifically, we consider an equimolar Li/M cation lattice and the Hamiltonian given in Eq. 1 and vary the relative contribution of $+J_1$ and $+J_2$ in Monte Carlo simulations according to:

$$\lambda = \frac{J_2}{J_1+J_2}$$

while maintaining the total interaction strength and macroscopic composition. We use this axis rather than the conventional $J_2/J_1$ axis such that the endpoints correspond to pure $J_1$ and pure $J_2$ character.

Fig. 3 shows the continuous evolution of the diffuse scattering shape across this interaction landscape. The $J_1$-only limit produces the cross-like scattering often described as tetrahedral-type.[5,20,21] Introducing $J_2$ transforms this pattern continuously through diamond- and rounded-square-shaped distributions resembling those reported experimentally for $Li_{1.2}Zr_{0.4}Mn_{0.4}O_2$ and $Li_{1.2}Ti_{0.4}Mn_{0.4}O_2$, respectively.[3] At $J_2/J_1$ = ½ (corresponding to λ =1/3), the Fourier-transformed pair interaction can be written as:

$$\mathbf{q} \sim [\cos(\pi h) + \cos(\pi k) + \cos(\pi l)]^2$$

whose minima reproduce the idealized octahedral surface as derived by De Ridder et al..[21] Further increasing $J_2$ produces cube-like scattering in the $J_2$-only limit. The conventional polyhedral description therefore represents regions of a continuous $J_1$-$J_2$ interaction landscape, rather than three structurally independent types (further discussion is provided in Supplementary Discussion 1). A related connection between diffuse-scattering morphologies through mathematical combinations of the De Ridder surfaces was performed by Li et al..[5]

To determine what these patterns imply in real space, we calculated the complete cation-composition distributions of the tetrahedral, octahedral and cubic environments (Fig. 3b-d). Relative to the random reference, the $J_1$-only structure contains an excess of balanced $Li_2M_2$ tetrahedra. This population decreases as $J_1$ is replaced by $J_2$, and the tetrahedral distribution approaches the random limit. Conversely, increasing $J_2$ strongly increases the relative population of $Li_4M_4$ cubes. The polyhedral composition distributions are not, however, exclusive fingerprints of the interaction that produced them. The cubic distribution in the $J_1$-only structure is broader than random and contains enhanced $Li_8$ and $M_8$ populations, arising indirectly from the correlations required to satisfy the $J_1$ interactions on the fcc lattice. Octahedral distributions are still less specific because they contain both first- and second-neighbour cation pairs. Both $J_1$ and $J_2$ can therefore favour balanced $Li_3M_3$ octahedra, explain-ing why their composition distribution does not identify a unique "octahedral" point in the interaction landscape. The corresponding Warren-Cowley relationships[39] and neighbour-shell derivations are given in Supplementary Methods 1 and Supplementary Discussion 2.

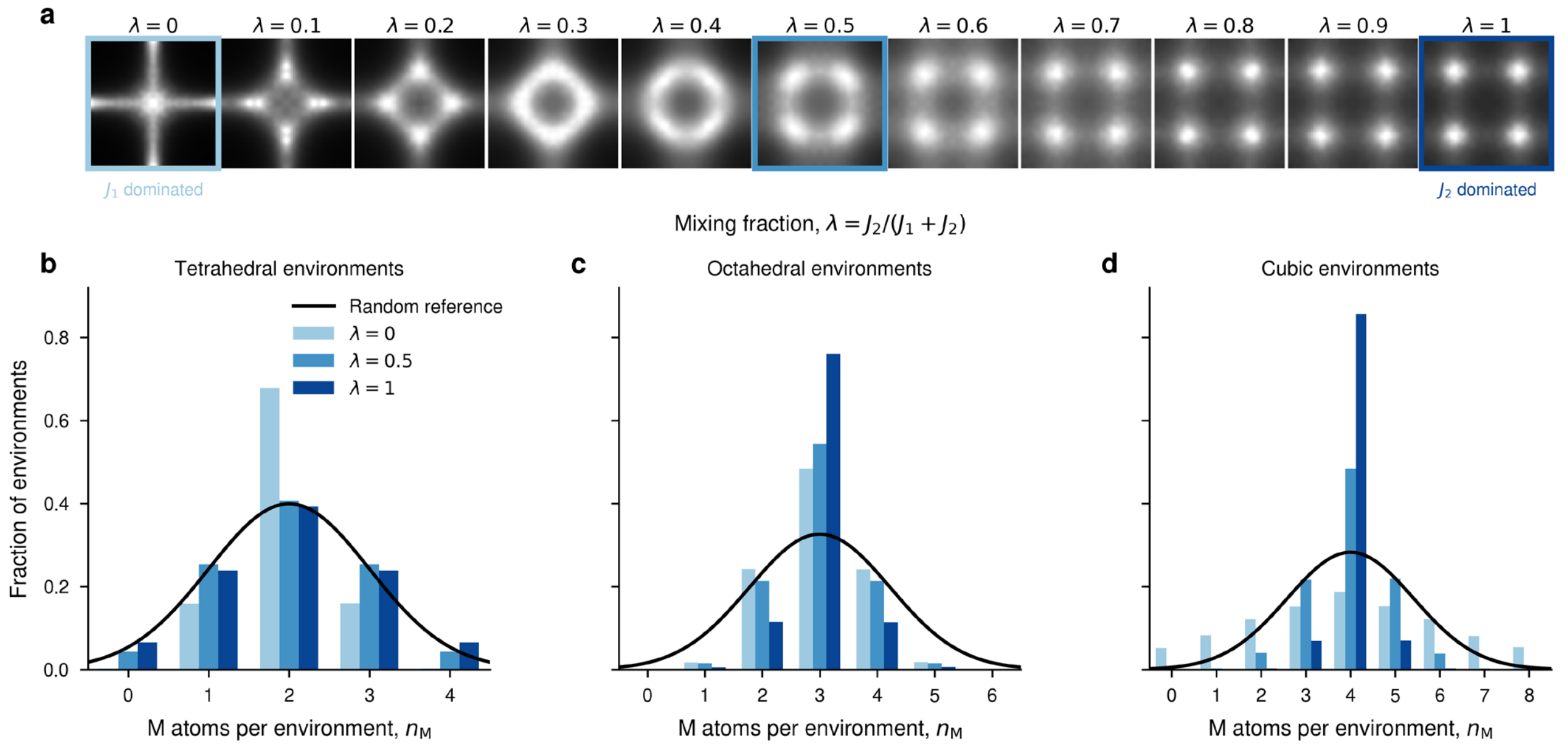

**Fig. 3 | Evolution of diffuse scattering and local cation environments across the pair-interaction landscape.** a, Calculated *hk*0 diffuse-scattering sections for an equimolar $LiMO_2$ rocksalt model as the mixing parameter $\lambda = J_2/(J_1 + J_2)$ increases from 0 to 1. Surfaces are plotted in the region bound by the reciprocal lattice points (0, 0, 0), (2, 0, 0), (0, 2, 0) and (2, 2, 0). b–d, Normalised composition distributions of tetrahedral (b), octahedral (c) and cubic (d) cation environments for λ = 0, 0.5 and 1. The horizontal axes give the number of M atoms within each environment. Black curves show the random-mixing reference. Increasing λ shifts the strongest compositional constraint from tetrahedral towards cubic environments, while balanced octahedral environments remain enriched throughout the sweep.

These comparisons provide a reference for interpreting the LTOF correlations, and when connecting diffuse scattering to lithium transport. Positive $J_1$ suppresses Li–Li first-neighbour pairs and therefore reduces the probability that all four vertices surrounding a tetrahedral migration site are occupied by Li. By contrast, the $J_2$-dominated reference structures retain an approximately random first-neighbour distribution and consequently a larger population of 0-TM environments than the $J_1$-dominated structures. Within this positive-interaction sweep, the diffuse scattering morphology varies with the relative contributions of $J_1$ and $J_2$, whereas the strength of the correlations – and hence their quantitative effect on the 0-TM network – also depends on the absolute interaction energies relative to temperature and the overall composition. Diffuse-scattering morphology alone, within this $+J_1$, $+J_2$ regime, therefore, cannot establish the extent to which short-range order affects lithium transport.

LTOF differs from these reference structures through its experimentally observed preference for like-cation first neighbours. In the minimal interaction model, this is reproduced in the $J_1/T < 0$, $J_2/T > 0$ regime. This regime is particularly favourable for increasing the population of 0-TM environments while limiting segregation. Positive $J_1$ favours unlike-cation first neighbours and generally suppresses $Li_4$ tetrahedra relative to random mixing. When both $J_1$ and $J_2$ are negative, like-cation preferences at both neighbour distances instead promote Li-rich and transition-metal-rich domains and, for sufficiently strong interactions, phase separation. Combining negative $J_1$ with positive $J_2$ promotes local Li enrichment while counteracting segregation through preferential unlike-cation pairing along ⟨100⟩. The correlations observed in LTOF therefore offer a route to increasing the 0-TM population without simply driving macroscopic cation segregation.

The positive first-neighbour cation correlation is consistent with coupled cation–anion short-range order, in which preferential Li coordination around F is accompanied by complementary Ti enrichment around O.[10] This partitioning increases the populations of both Li–Li and Ti–Ti first-neighbour pair populations at the expense of Li–Ti pairs. Because O and F have similar electron-scattering factors, their occupational correlations cannot be resolved directly. The observed cation order is therefore consistent with coupled Li–F/Ti–O preferences and local charge balance,

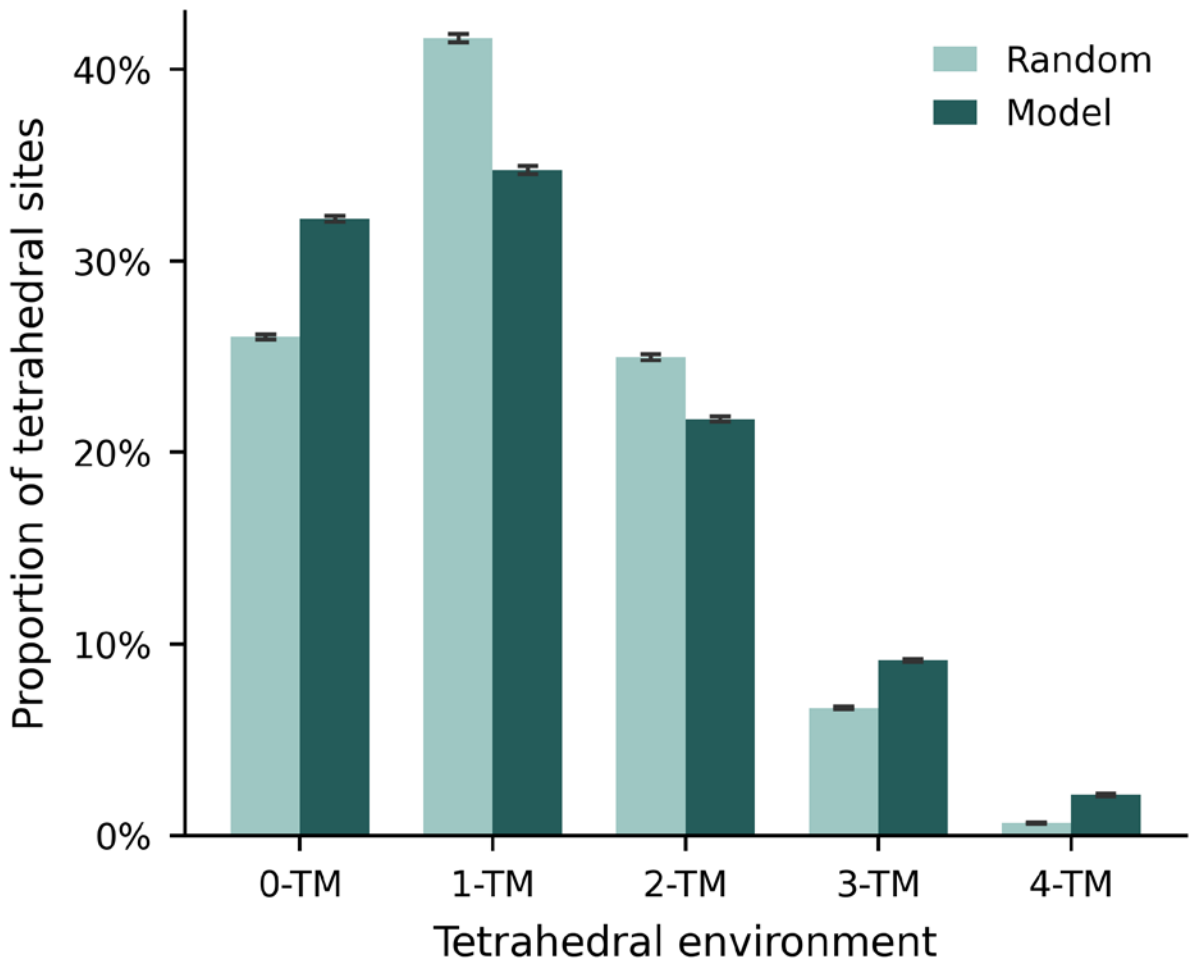

**Fig. 4 | Population of 0-TM tetrahedral sites.** Proportions of *n*-TM tetrahedral environments in the pair-interaction model compared with random cation mixing at the same composition. Here, *n*-TM denotes a tetrahedral site sharing faces with *n* transition-metal-occupied and 4 - *n* Li-occupied octahedral sites.

but the relative contributions of these interactions cannot be determined from these measurements.

The negative second-neighbour correlation at (1,0,0) can be understood from the accompanying displacive correlations. Although coupled Li–F/Ti–O ordering might also be expected to generate some like-cation second-neighbour pairs, the experimentally observed correlation is instead strongly negative. At the nearest cation–anion vector (½,0,0), the anions displace towards Ti and away from Li, accommodating the shorter preferred Ti–X and longer Li–X distances (Fig. 2a).[36] For two cations separated by (1,0,0), this displacement is most readily accommodated by an unlike Ti–X–Li configuration, in which the intervening anion can move towards Ti while simultaneously moving away from Li. The tendency of shared-atom relaxation to distinguish first- and second-neighbour pairs has a precedent in interstitial Pd–B, where elastic interactions favour first-neighbour B pairs but disfavour second-neighbour pairs.[40] The observed negative second-neighbour correlation is therefore consistent with strain accommodation favouring Ti–X–Li configurations. This provides a plausible interpretation of the coupled occupational and displacive correlations, although establishing their energetic origin requires further investigation.

## Implications for Li-migration

The like-cation first-neighbour preference observed in LTOF suggests that its short-range order can enrich local environments favourable for lithium migration. We assessed the 0-TM network at three levels: the population of 0-TM tetrahedra, the connectivity of Li sites through 0-TM channels, and the directness of paths through that network. Values were averaged over 100 large-box configurations with the standard deviation providing an error analysis of the calculated values.

The 3D-ΔPDF-informed model contains 32.20(16)% 0-TM tetrahedra, compared with 26.04(14)% for random cation mixing at the same composition (Fig. 4). All six edges of a cation tetrahedron join first-neighbour cation sites; the preference for like-cation first neighbours therefore increases the occurrence of compositionally enriched tetrahedra. The resulting redistribution increases the $Li_4$ population by 6.2 percentage points relative to random mixing. Our results support and realise the prediction by Liu et al. that favouring like-cation first neighbours and unlike-cation second neighbours can increase the 0-TM population above the random limit.[9] As these population values are highly dependent on the strength of $J$, Supplementary Fig. 14 shows robustness of the higher-than-random 0-TM populations for weaker $J_1$ and $J_2$.

We next evaluated whether these additional 0-TM channels join the existing network. Following the percolation approach introduced for rocksalt cathodes by Urban *et al.*, we connected neighbouring octahedral Li sites when at least one intervening tetrahedral channel was 0-TM. The fraction of Li in the largest system-spanning network rises from 92.75(22)% for random mixing to 97.17(14)% for the 3D-ΔPDF-informed model. Thus, the increased $Li_4$ population modestly enlarges an already extensive network. Because Ti is already 4+, these results quantify network topology rather than demonstrating electrochemical accessibility or lithium transport in $Li_{2.5}TiO_3F_{0.5}$.

The model network has approximately 16% lower tortuosity than the random-cation network, indicating more direct paths through its 0-TM channels. Because the model also contains more 0-TM tetrahedra, this comparison cannot establish how much of the reduction arises from their spatial arrangement rather than their increased population. The short-range order nevertheless offers a possible geometric explanation. Net migration along ⟨100⟩ requires a zigzag sequence of octahedral-site hops, for example $(0,0,0) \rightarrow (\frac{1}{2},\frac{1}{2},0) \rightarrow (1,0,0) \rightarrow (\frac{3}{2},\frac{1}{2},0)$, whereas successive hops along ⟨110⟩ can form a straight path, $(0,0,0) \rightarrow (\frac{1}{2},\frac{1}{2},0) \rightarrow (1,1,0)$. The $J_1/T < 0$, $J_2/T > 0$ correlations may favour arrangements of 0-TM channels supporting the latter paths, but their contribution to the measured tortuosity reduction cannot be separated from the change in channel population here. In the limiting case where a negative four-body interaction selects spinel-like order, the 0-TM tetrahedra form a pyrochlore-like corner-sharing network with effectively straight migration paths. The experimentally informed model therefore connects the observed short-range order with an increased population of 0-TM environments, a larger connected Li network and more direct paths through that network.

Short-range order not only influences the cation distribution, affecting the 0-TM population and connectivity, but also the local atomic displacements. Local distortions around migration sites modify the electrostatic and geometric environments and, in turn, the activation energy for diffusion. Recent studies have demonstrated that these distortions can enable transport through 1-TM sites, extending the accessible diffusion network beyond the consideration of 0-TM sites alone.[12,13] The correlated displacements resolved in LTOF may similarly modify local migration barriers, although the direction and magnitude of this effect remain to be established. By resolving these displacements together with occupational correlations, electron 3D-ΔPDF provides experimental constraints for atomistic models that can assess how both contributions influence lithium migration.

In summary, we have demonstrated the application of electron 3D-ΔPDF analysis to cation-disordered rocksalt materials, resolving both occupational and displacive short-range correlations. In $Li_{2.5}TiO_3F_{0.5}$, we identify atypical cation ordering that increases the population and connectivity of 0-TM migration sites, with substantial correlated atomic displacements. Representative models constrained by these correlations exhibit an above-random population of 0-TM environments, a larger connected lithium network and reduced geometric tortuosity. The analysis also reveals directional anion and cation displacements whose inclusion substantially improves agreement with independent powder-PDF measurements, establishing local relaxation as an essential component of the structural description. These displacements may additionally modify lithium migration barriers, although their effect on transport remains to be determined. Together, the findings demonstrate how electron 3D-ΔPDF can move beyond qualitative identification of disorder to constrain atomistic models incorporating both chemical order and local distortions, providing an experimental foundation for understanding their combined influence on lithium transport in disordered rocksalt materials.

## Methods

### Monte Carlo simulations

Periodic disordered-rocksalt models were constructed from cubic conventional cells with a lattice parameter of 4.10 Å. For the $LiMO_2$ modelling, a 20 × 20 × 20 supercell contained 32,000 cation sites and 32,000 anion sites. Cation sites were randomly occupied by Li and M in a 1:1 ratio, where M had the scattering factor of Fe. For modelling the experimental data, a 21 × 21 × 21 supercell contained 37,044 cation sites and 37,044 anion sites, with Li and Ti occupying the cation sublattice in fractions of 5/7 and 2/7, respectively. All anion sites were occupied by O, and periodic boundary conditions were applied in all three dimensions.

Short-range cation order was generated using fixed-composition Metropolis Monte Carlo sampling. The energy of the system was evaluated by the Hamiltonian stated in the main text. Trial moves exchanged one randomly selected M atom and one randomly selected Li atom with the change in the energy of the system calculated locally by:

$$\frac{\Delta\mathcal{H}}{k_B T} = J_1\big([z_1^{Ti}(b) - I_1(a,b) - z_1^{Ti}(a)]\big) + J_2\big([z_2^{Ti}(b) - I_2(a,b) - z_2^{Ti}(a)]\big)$$

Where $z_s^{Ti}(i)$ is the number of Ti neighbours of site *I* in shell *s*, and $I_s(a,b)$ is 1 when *a* and *b* belong to the shell *s* and 0 otherwise. Moves with $\Delta\mathcal{H} \leq 0$ were accepted, whereas moves with $\Delta\mathcal{H} > 0$ were accepted with the probability $\exp\left[\frac{-\Delta\mathcal{H}}{k_b T}\right]$.

For the $LiMO_2$ mixing sweep, $J_1 = (1 - \lambda)J_{total}$ and $J_2 = \lambda J_{total}$ with $J_{total} = 2.32$ eV and $k_B T = 1$. Each simulation comprised 2,000,000 attempted exchanges, corresponding to 62.5 attempts per cation. One hundred configurations were sampled evenly during the final 1,000,000 attempts. For the Li/Ti model, $J_1/T = –0.5$, $J_2/T = +0.5$. Twenty exchanges were attempted per cation, and the final occupational configuration was retained.

For Fig. 4, an ensemble of 100 independent Li/Ti occupational models was generated using the same supercell, composition, Hamiltonian, and Metropolis sampling procedure. Each model was initialized with a separately randomized cation arrangement and subjected to 20 attempted exchanges per cation site, corresponding to 740,880 attempted exchanges per trajectory. A reference ensemble of 100 independently randomized configurations was generated at the same fixed composition, without subsequent MC sampling. Tetrahedral-site populations and connectivity were evaluated from the occupational configurations on the ideal rocksalt lattice. Reported ensemble averages were calculated across the 100 configurations, with error bars representing the sample standard deviation.

**Static displacement modelling**

Using a local pair-based displacement model, static displacements were assigned to the Monte Carlo configurations produced by $J_1/T$, $J_2/T = –0.5, +0.5$. Each oxygen atom received a 0.1 Å vector contribution towards each neighbouring Ti along ⟨100⟩; contributions from neighbouring Ti were summed vectorially. For every nearest-neighbour Ti-Ti pair, each Ti received a 0.025 Å contribution away from one another along the interatomic ⟨110⟩ direction. For every second-neighbour unlike-cation pair, both Li and Ti received a 0.075 Å contribution directed away from the other ion along ⟨100⟩.

**Diffuse-scattering calculations**

Diffuse scattering was calculated from the big-box models using Scatty (version dated 14 October 2022).[41] Electron scattering factors were used for the Li/Ti model compared with the experimental data, while neutron scattering factors were used for the conventional rocksalt simulations in Fig. 3 to avoid visual effects from the scattering-factor drop off.

**Synthesis**

The synthesis employed here was adapted from a route reported for $Li_3TiO_3F$.[42] To achieve the target composition, stoichiometric amounts of $Li_2CO_3$ (544.3 mg, Fluka, >98%), anatase $TiO_2$ (588.6 mg, Sigma-Aldrich, >99%), and LiF (95.3 mg, Sigma-Aldrich, >99%), were thoroughly ground using a mortar and pestle. The resulting powder was pressed into 20 mm diameter pellets at 5T for 5 min. The pellets were subsequently placed in a tube furnace and heated at 5 °C/min to 800 C, held for 24 h, and then quenched and stored in air.

**Powder X-ray diffraction**

Powder X-ray diffraction data were collected in-house on a Rigaku SmartLab diffractometer equipped with an XSPA-400 ER Seamless Pixel Array 2D detector. The diffractometer employed a rotating anode source with a Rigaku Kα1 Johansson monochromator to generate Cu *Kα*1 radiation (λ = 1.54056 Å). The measurement was performed in Bragg-Brentano mode and 2.5° Soller slits were used on both the incident and detector sides. Prior to measurement, the sample was ground with a mortar and pestle and placed on a zero-background Si sample holder. Measurements were performed over a $2\theta$ range of 10°-100° at a scan rate of 3°/min. The pattern revealed the target composition with ~93 % purity and a small amount (~7%) of monoclinic $\beta$-$Li_2TiO_3$ as a secondary phase. Rietveld refinement was performed in TOPAS.[43–45] A background polynomial, axial peak asymmetry correction, sample height correction, unit cell parameters, scale factors and isotropic temperature factors were allowed to refine.

**Powder total scattering and PDF analysis**

Total scattering measurements were performed at beamline ID31 at the European Synchrotron Radiation Facility (ESRF). The sample powders were loaded into cylindrical slots (approx. 1 mm thickness) held between Kapton windows in a high-throughput sample holder. Data on each sample were measured in transmission geometry with an incident X-ray energy of 75.050 keV (λ = 0.16520 A). Measured intensities were collected using a Pilatus CdTe 2M detector (1679×1475 pixels, 172×172 $\mu m^2$ each) positioned with the incident beam in the corner of the detector. The sample-to-detector distance was approximately 0.3 m. Multiple measurements of the empty well with polyimide windows were measured, and summed together for improved statistics, for the background subtraction.

NIST SRM 660b ($LaB_6$) was used for geometry calibration. Integration was performed with the software pyFAI including flat-field, geometry, solid-angle, and polarization corrections. Invalid pixels were masked, and the sigma clipping integration algorithm was used to automask further azimuthal outliers. The summed background intensities were scaled respectively to each of the reference and sample measurements and subtracted. Rietveld refinements were performed using TOPAS v7[43–45] to fit the structure with the reference lattice parameter fixed to the measurement to determine the instrumental profile contribution and correct for offset errors including parallax.[46]

PDF data were processed in an automated way using PDFgetX3[47] using a Lorch modification function to suppress termination effects and contributions from high frequency noise. Total scattering data out to 25 $Å^{-1}$ Q was used in the Fourier transform.

PDF refinements of the LaB6 NIST SRM 660b and Silicon NIST SRM 640c reference standards were performed using PDFgui[48] to verify the quality of the real-space data. The resolution parameters refined to values of Qdamp = 0.008 $Å^{-1}$ and Qbroad = 0.013 $Å^{-1}$.

Rietveld-like refinements using the PDF data were performed in TOPAS-Academic v8.[43–45] A single temperature factor was refined per average structure phase. 10x10x10 unit cell large box models were truncated from the centre of 21x21x21 unit cell models produced by the Monte Carlo simulations and subsequent displacement modelling. For these, one temperature factor was used per atom type. Cell parameters were refined as well as the scale factors where the scale factor of the impurity phase was constrained to 7 wt% (converted to 7.76 mol% for PDF refinement in TOPAS).

**STEM-EDX**

The sample was suspended in ethanol (99.9%) and sonicated for 5 min. Afterwards, 3 drops of 5 µl were drop casted on a TEM grid (ultrathin carbon film on lacey carbon support film, 400-mesh Cu, Plano GmbH).

STEM-EDX was performed on two crystallites after the electron diffraction experiments using a ThermoFisher Scientific Talos F200X FEI S/TEM equipped with an X-FEG operated at 200 keV and equipped with a Bruker ChemiSTEM system. This revealed Ti:O:F percentages of 23:67:10; lithium cannot be quantified by STEM-EDX but can be informed by the other elements through charge and occupancy balancing. This gives an F/Ti ratio of 0.43, modestly below the nominal value of 0.5 although light-element analysis carries appreciable uncertainty. The diffuse scattering analysis was carried out assuming a $Li_{2.5}TiO_3F_{0.5}$ composition and Supplementary Fig. 9 shows the model is robust to these subtle composition differences.

**Electron diffraction measurements**

Electron diffraction was performed using a ThermoFisher Scientific Talos F200X FEI S/TEM equipped with a XFEG operated at 200 keV. The diffraction was acquired with a Merlin 4R hybrid pixel detector from Quantum detector.

To acquire the diffraction at various zone-axes, a double-tilt holder was used. The instrument was operated at microprobe mode with a convergence angle of 0.6 mrad, corresponding to an approximated beam diameter of 5.14 nm, neglecting chromatic and higher order aberrations. The tilt was adjusted to obtain a (near) zone-axis orientation for several different crystals. Each diffraction pattern was recorded with a frame time of 4 ms on 256x256 pixels with a 12-bit as bit depth and a reciprocal space resolution of 0.016935 $Å^{-1}$/pixel. To improve signal-to-noise ratio, the sum over several diffraction patterns from each crystal was used for further analysis. To obtain the orientation of each pattern, the orientation was matched with SingleCrystal software (version 4.1.9, SingleCrystal Software Ltd., Oxfordshire, UK).

The static frame rotation experiment was performed using the same microscope but operating it in TEM mode and using 512x512 pixels of the Merlin detector with a 24-bit bit depth. The reciprocal space resolution was 0.008157 $Å^{-1}$/pixel. A selected area aperture of 10 µm was used, limiting the area from which the diffraction comes to have a diameter of ~220 nm. To allow for higher tilt angles, a tomography holder was used. Reciprocal space slices for every 1° between –45° and +45° were recorded for the main dataset. For the larger secondary crystallite, the bounds were –50° and +50°. The crystallite dimensions in the plane of the grid are ~100 × 200 nm and ~200 × 200 nm, respectively. For each tilt 200 diffraction patterns, each with a frame time of 100 ms, were recorded. The sum of these 200 patterns was used for all further processing steps. the

**Data processing**

The TIFF files recording the diffraction frames were processed with PETS2. A reflection diameter of 20 pixels was used for the peak search. A reciprocal space volume was constructed out to 2.91 $Å^{-1}$ ($Q_{max}$ = 18.3 $Å^{-1}$) with a pixel size of 0.0081 $Å^{-1}$ including the built-in background subtraction.

The scattering volume was then symmetrized based on the $m\bar{3}m$ Laue group. During symmetrisation, outlier rejection was applied to symmetry related voxels. The outlier rejection algorithm follows the approach outlined by Blessing with the parameters set to $c_1 = c_2 = c_4 = 0$ and $c_3 = 12$.[49]

Following symmetrisation, the Bragg peaks were removed and missing values interpolated using a variation of the punch-and-fill method.[50] The process is summarized in Supplementary Methods 2.

The 3D-ΔPDF was obtained directly from the isolated diffuse scattering by applying the fast Fourier transform as implemented in NumPy.[51]

To deconvolute the occupational disorder on the cation sites from the displacive disorder a box integration approach was employed. The resulting map where features of the 3D-ΔPDF have been integrated holds information only on the occupational correlations (see Supplementary Discussion 3).[35,52]

**Acknowledgments**

We thank the Danish National Research Foundation (Grant No. DNRF189) through the Center for Sustainable Energy Materials (CENSEMAT) for funding this research. The numerical results presented in this work were obtained at the Centre for Scientific Computing, Aarhus with the help of the Grendel cluster.

We acknowledge the European Synchrotron Radiation Facility (ESRF) for provision of synchrotron radiation facilities and Momentum Transfer for facilitating the measurements. Jakub Drnec is thanked for assistance and support in using beamline ID31. The measurement setup was developed with funding from the European Union's Horizon 2020 research and innovation program under the STREAMLINE project (grant agreement ID 870313).

We thank Prof. John S. O. Evans and Nikolaj Roth for inspiring discussions regarding short-range order in fcc materials.

**Data availability**

The experimental electron diffraction, diffuse-scattering and 3D-ΔPDF datasets, together with the structural models and supporting data generated in this study, are available to editors and reviewers through Zenodo at the private link provided in the submission portal. The numerical data underlying the main figures are also available in the associated GitHub repository (https://github.com/samyulethompson/LiTiOFDiffuseScattering).

**Code availability**

Custom Python scripts used for Monte Carlo simulations, structural analysis and figure generation are available at https://github.com/samyulethompson/LiTiOFDiffuseScattering. The version of the code associated with this publication will be archived on Zenodo at publication. Scripts relating to the processing of the diffuse scattering data can be obtained from the authors upon request.

**Author Contributions**

SYT | conceptualisation, data curation, formal analysis, investigation, methodology, software, visualisation, writing – original draft, writing – review & editing; KAHS | conceptualisation, data curation, formal analysis, investigation, methodology, software, validation, writing – original draft, writing – review & editing; NZ | formal analysis, investigation, writing – original draft, writing – review & editing; RK | data curation, formal analysis, investigation, software, validation, writing – original draft, writing – review & editing; EMS | conceptualisation, formal analysis, methodology, writing – review & editing; SW | methodology, software, writing – review & editing; NU | methodology, software, writing – review & editing; JS | funding acquisition, supervision, writing – review & editing; EDB | conceptualisation, funding acquisition, resources, supervision, writing – review & editing; GKHM | funding acquisition, supervision, writing – review & editing; BBI | conceptualisation, funding acquisition, supervision, writing – review & editing; DBR | conceptualisation, funding acquisition, project administration, resources, supervision, writing – original draft, writing – review & editing.

## Supplementary Information

**Supplementary Methods 1**: local-environment and short-range-order analysis

Local cation environments were analysed from the Monte Carlo configurations using a custom Python script. Each tetrahedra, octahedra and cubes were counted once according to the position of its geometric centre. For a configuration containing $N_{cat}$ cation sites, the analysis generated $2N_{cat}$ tetrahedra, $N_{cat}$ octahedra and $N_{cat}$ cubes.

For each environment, the number of M cations, $n_M$, was counted and grouped into N+1 discrete compositions. Each histogram was normalized by the total number of environments to achieve a probability distribution.

The random reference distributions were calculated using the exact binomial expression

$$P_{\mathrm{random}}(n_{\mathrm{M}}) = \binom{N}{n_{\mathrm{M}}} x_{\mathrm{M}}^{n_{\mathrm{M}}} (1 - x_{\mathrm{M}})^{N-n_{\mathrm{M}}}$$

where $x_M$ = 0.5 for the equimolar model.

The width of each polyhedral distribution was quantified relative to its random reference using

$$R_{\mathrm{P}} = \frac{\mathrm{Var}(n_{\mathrm{M}})}{N x_{\mathrm{M}}(1 - x_{\mathrm{M}})}$$

Where P denotes the tetrahedral, octahedral or cubic environment. Values of $R_P$ =1, $R_P$ < 1 and $R_P$ > 1 indicate random, narrower and broader composition distributions, respectively. The relationship between these distribution widths and the Warren-Cowley parameters are derived in Supplementary Discussion 2.

Warren-Cowley parameters were calculated for cation pairs up to the sixth radial neighbour shell of the fcc lattice:

$$\alpha_r = 1 - \frac{P(\mathrm{M}|\mathrm{Li}; r)}{x_{\mathrm{M}}}$$

Where P(M|Li; *r*) is the probability that a site in neighbour shell *r* around Li is occupied by M. Under this convention, $\alpha_r = 0$ represents random mixing, $\alpha_r < 0$ indicates a preference for unlike Li-M pairs and $\alpha_r > 0$ indicates enhanced like-cation pairing.

The population of 0-TM migration environments was obtained directly from the tetrahedral composition distribution. A tetrahedron was classified as 0-TM when all four surrounding cation sites were occupied by Li, corresponding to $n_{\mathrm{M}} = 0$.

**Supplementary Methods 2**: Punch-and-fill approach for electron scattering data

While similar to single crystal X-ray scattering in many aspects three-dimensional ED (3D-ED) differs in ways which have important implications for the punch-and-fill approach used in the 3D-ΔPDF data reduction. One of the primary differences is that unless the electron diffraction instrument is equipped with an energy filter, the inelastic scattering of electrons will lead to a radial broadening of the Bragg peaks. This means, that in cases where the diffuse scattering overlaps with or is concentrated in regions near the Bragg peaks, a conventional spherical punch shape will be non-ideal as it will either be too large, resulting in removal of diffuse scattering, or too small, leading to incomplete removal of the Bragg intensities. It is therefore necessary to use a punching method which handles anisotropic peak shapes. One approach could be the KAREN algorithm where Bragg peaks are identified as outliers using traditional signal processing methods based on the Hampel filter.[1] However, the KAREN algorithm does not utilize the information provided by the unit cell and orientation matrix. For all voxels, including voxels not centred at integer Miller indices, it evaluates whether the voxel is an outlier with respect to the neighbouring voxels in a defined window region. As such, sharp diffuse scattering resulting from the presence of long-range correlations may inadvertently be marked as outliers.

A way to avoid this, is to use the information provided by the unit cell parameters and orientation matrix, by only searching for Bragg intensities in a selected volume centred at integer Miller indices. The Bragg intensity search can be performed in various ways. The approach used here is inspired by the seed-skewness method initially developed for integrating single crystal Bragg intensities on 2D imaging plate detectors.[2,3]

Initially a 3D box is centred at the ideal Bragg peak position. The box size is chosen sufficiently large to encompass both the Bragg peak and the surrounding background. Before performing any computations, the intensities within the box are shifted by the absolute value of the largest negative value. As such, the box will only contain positive intensities. The average background intensity within the box, $\langle B \rangle$, and the background intensity standard deviation, $\sigma(B)$, is estimated from the intensities along the faces of the box. An initial seed for the Bragg peak mask is determined as all voxels with intensities higher than $\langle B \rangle + x_{\mathrm{crit}}\sigma(B)$. $x_{\mathrm{crit}}$ here represents a tuneable parameter where a small value will lead to a larger initial mask (seed) and fewer iterations in the dynamic mask dilation – however, the mask will be more sensitive to the presence of diffuse scattering near the boundaries of the box. A larger value of $x_{\mathrm{crit}}$ will on the other hand result in a smaller initial mask (seed) and require more iterations in the dynamic mask dilation while providing a seed and final mask that is less sensitive to diffuse scattering near the box boundaries. For the data presented in this paper a value of $x_{\mathrm{crit}} = 15$ was used.

Having defined the seed (i.e. the initial Bragg peak mask) a new set of background voxels can be defined as the voxels in the box, not masked by the seed. The skewness of the intensity distribution of these background voxels can be calculated as:

$$\mu_3 = \frac{1}{N}\sum_{i=1}^{N}(I_i - \langle I \rangle)^3$$

The neighbour voxels to the seed are searched for the voxel with the highest intensity which is then added to a new temporary peak mask. Using the temporary mask a new estimate for the skewness of the background intensity distribution can be made, $\mu_3^{\mathrm{temp}}$. If $\mu_3^{\mathrm{temp}} < \mu_3$ and $\frac{|\mu_3^{\mathrm{temp}} - \mu_3|}{\mu_3} > 10^{-3}$ the mask dilation is accepted and $\mu_3^{\mathrm{temp}}$ replaces $\mu_3$. Iterative attempts at mask dilation are then performed until the process converges in $\mu_3$. Assuming Poisson statistics for the pixel intensities convergence of $\mu_3$ to a non-zero value should be achieved. In the specific case of high expected average counts, the total background intensity should approach a normal distribution with $\mu_3 = 0$. However, given the presence of diffuse scattering in our data, a skewness of zero is unlikely to be achieved by removal of the Bragg intensities.

It is an inherent issue for the seed-skewness method that the background tends obscure the tails of the Bragg peak.[3] For the present application this implies that the size of the Bragg peak mask is underestimated leading to a small amount of Bragg intensity bleeding out into the background voxels. A simple solution to this problem is to define a parameter, $N_{\mathrm{dil}}$, which specifies the number of times a given mask should be dilated by including all nearest neighbour voxels in the mask. For the data presented here we used $N_{\mathrm{dil}} = 1$ thus performing only one dilation step of the seed-skewness masks. As a final step, any internal holes in the masks are filled using the binary_fill_holes algorithm implemented in SciPy.[4]

The effect of the seed-skewness based punching algorithm is illustrated for the (200) and (220) reflections in Supplementary Fig. 1 and 2 respectively. In Supplementary Fig. 1a and 2a the scattering intensity within the box is shown as two-dimensional slices. Low intensities are shown in white while high intensities are shown in red. The intensity colour map has been overlaid with a semi-transparent black mask which has a transparency of $\alpha = 0$ for voxels assigned to the Bragg peak and $\alpha = 0.6$ for voxels assigned to the background. As such, background voxels appear greyed out in the figure. From visual inspection, the mask appears to nicely partition Bragg and background voxels for both the (200) and the (220) reflections. Supplementary Fig. 1b and 2b show histograms over the intensities of the background assigned voxels. In both cases, the distributions appear reasonably symmetric after removing the voxels contributing the most to the skewness of the distribution. Supplementary Fig. 1c and 2c show histograms over the intensities of the Bragg peak assigned voxels. As expected, these voxels correspond mainly to intensities contributing to the high intensity tail of the voxel intensity distribution (note that the histogram is cut off at intensities exceeding $2.5I_{\mathrm{max}}$ with $I_{\mathrm{max}}$ being the maximum intensity of the background assigned voxels). Finally, to verify that the seed-skewness algorithm acts as expected, it is helpful to visualize the skewness of the background distribution as a function of the iterative growth of the Bragg peak mask. Such plots are shown in Supplementary Fig. 1d and 2d for the (200) and the (220) reflections respectively. In both cases, the skewness is observed to decay to a constant value asymptotically as anticipated.

The seed-skewness punching method was applied to reflections with all-even Miller indices satisfying the requirement $\sqrt{h^2 + k^2 + l^2} < 6.5$. The remaining Bragg peaks were removed using spherical punches with a radius of 5 voxels.

The punched intensities were interpolated using the nearest-neighbour interpolation as implemented in SciPy.[4] Missing values beyond a resolution of 10.43 Å$^{-1}$ were filled with Gaussian noise resembling the average background intensity. The 3D-ΔPDF was obtained as the Fourier transform of the array containing the diffuse scattering using the FFT algorithm as implemented in NumPy.[5]

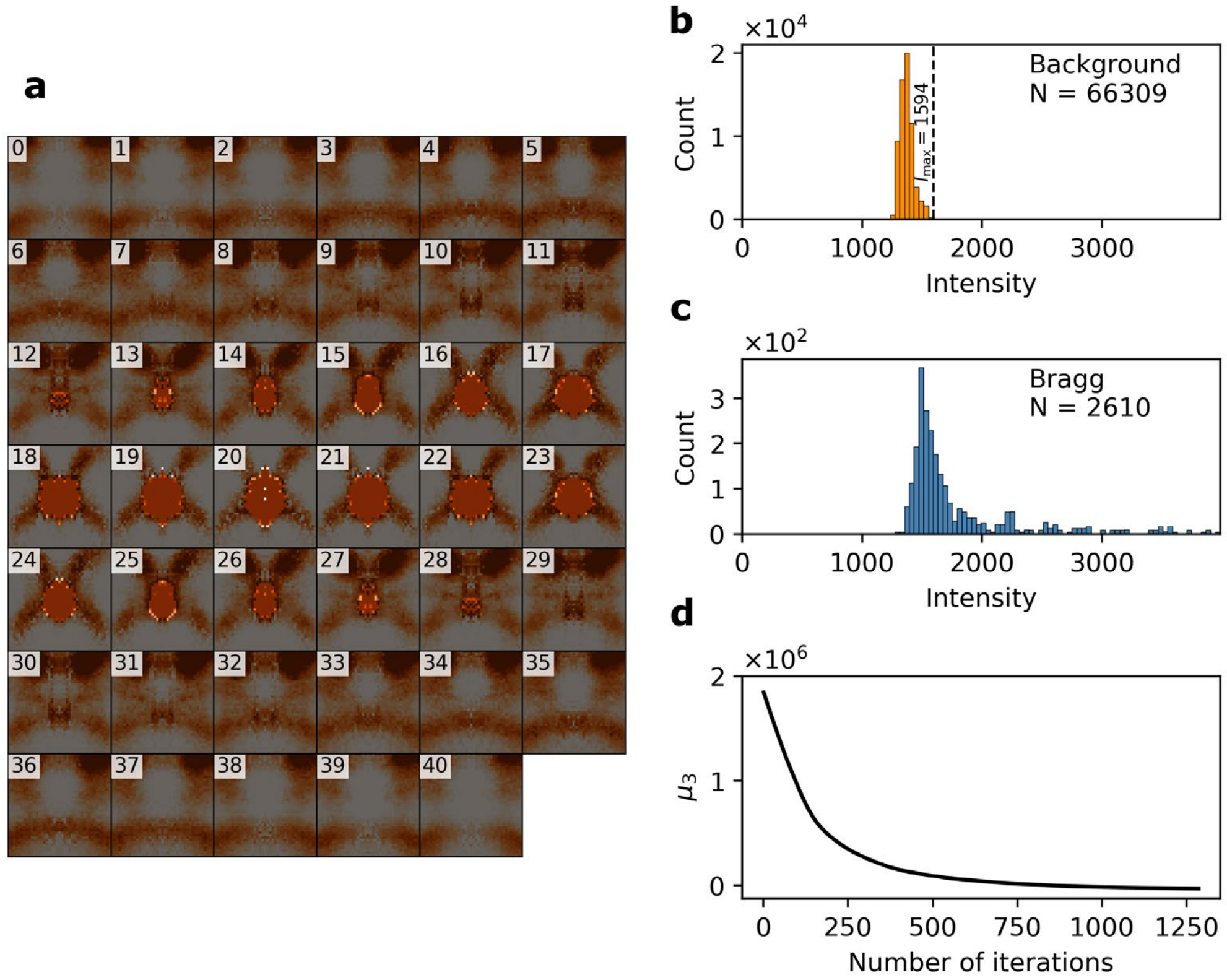

**Supplementary Fig. 1 | Evaluation of seed-skewness punching algorithm on the (200) reflection.** In **a** slices through the box containing the Bragg reflection are shown. High intensities are shown in red, while low intensities are shown in white. Greyed out regions correspond to regions belong to the background as determined using the seed-skewness algorithm. In **b** a histogram over the intensities of the background voxels is shown. The vertical dashed line marks the maximum intensity of the background voxels. The corresponding histogram over the intensities of the Bragg peak voxels is shown in **c**. The background skewness is plotted as a function of the seed-skewness iteration in **d**.

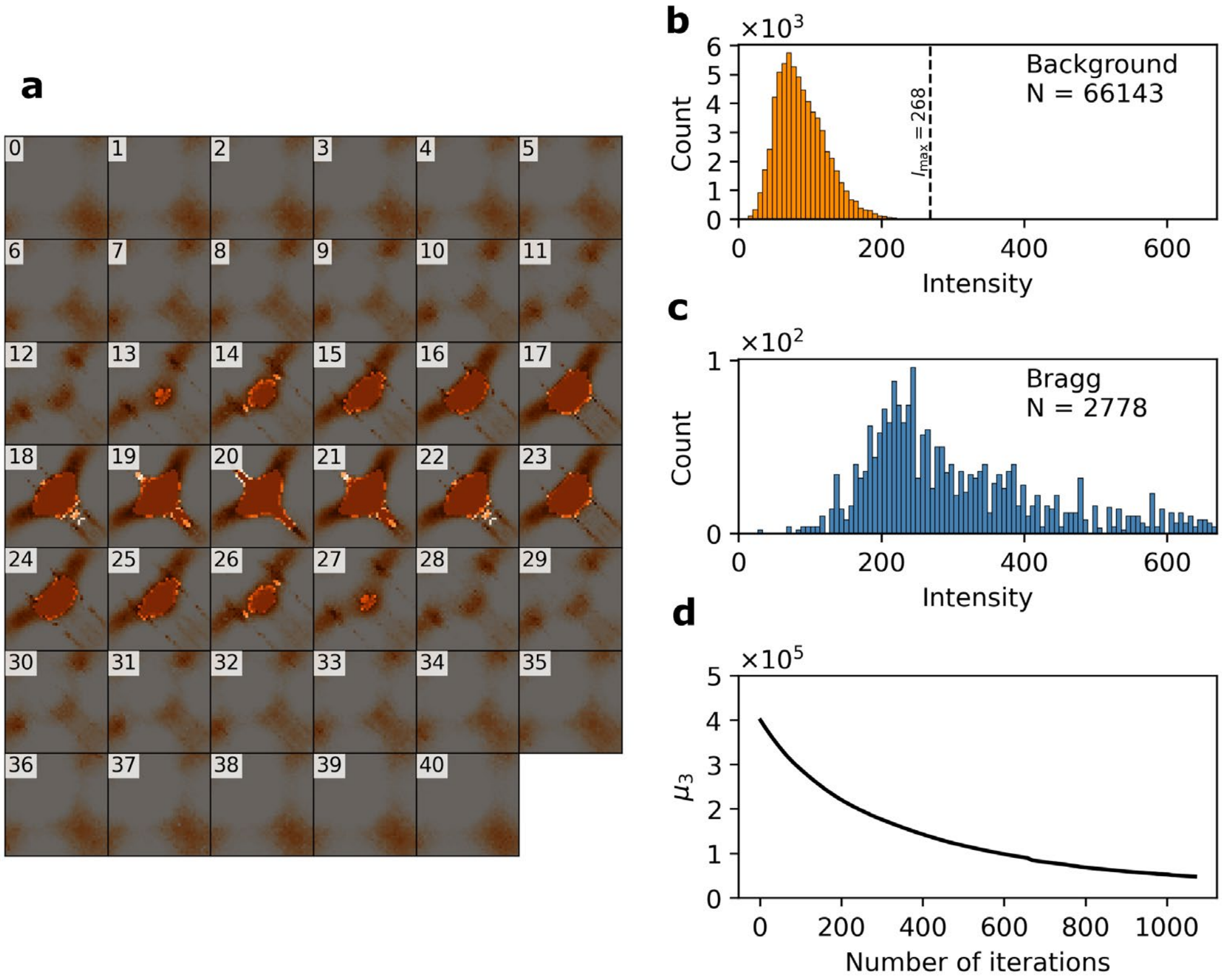

**Supplementary Fig. 2 | Evaluation of seed-skewness punching algorithm on the (220) reflection.** In **a** slices through the box containing the Bragg reflection are shown. High intensities are shown in red, while low intensities are shown in white. Greyed out regions correspond to regions belong to the background as determined using the seed-skewness algorithm. In **b** a histogram over the intensities of the background voxels is shown. The vertical dashed line marks the maximum intensity of the background voxels. The corresponding histogram over the intensities of the Bragg peak voxels is shown in **c**. The background skewness is plotted as a function of the seed-skewness iteration in **d**.

**Supplementary Discussion 1**: Relationship between pair interactions and De Ridder surfaces

The structured diffuse scattering observed in many disordered fcc materials has previously been described using the invariant-cluster model developed by De Ridder and co-workers.[1] In this model, the composition of a selected local cluster—such as a tetrahedron, octahedron or cube—is constrained to remain close to the bulk composition. For the equimolar model considered here, these constraints correspond to preferential formation of $Li_2M_2$ tetrahedra, $Li_3M_3$ octahedra or $Li_4M_4$ cubes. In this work, we study the effect of changing pair-interaction energies at the first- and second-neighbour cation distances and their effect on the diffuse scattering. This discussion will describe the relationship between this method and the De Ridder surfaces.

The tetrahedral De Ridder condition follows directly from the $J_1$-only limit. A tetrahedron on the fcc cation lattice contains six internal pairs, all of which are first-neighbour pairs. A positive $J_1$ therefore favours unlike cations along the six tetrahedral edges, suppressing compositionally extreme $Li_4$ and $M_4$ tetrahedra and favouring balanced $Li_2M_2$ environments. Because every pair within the tetrahedron is a first-neighbour pair, the $J_1$-only Hamiltonian is equivalent to penalizing deviations from a balanced tetrahedral composition. Consequently, the $\lambda = 0$, simulation produces diffuse intensity distributed around the published tetrahedral De Ridder surface.

The octahedral condition also follows from the pair model but requires both $J_1$ and $J_2$. An octahedron contains twelve first-neighbour pairs along its edges and three second-neighbour pairs connecting opposite vertices. When all overlapping octahedra in the fcc lattice are considered, each first-neighbour bond belongs to two octahedra, whereas each second-neighbour bond belongs to one. A penalty against deviations from a balanced $Li_3M_3$ octahedral composition therefore corresponds to first- and second-neighbour interaction strengths in the ratio $J_2/J_1 = ½$. In our mixing framework (SM1), this condition occurs at $\lambda = 1/3$. The scattering for this condition is shown in Fig. 1e. The pair-interaction model at $\lambda = 1/3$ is therefore equivalent to the local octahedral constraint considered by De Ridder *et al.* The resulting calculated scattering follows the ideal octahedral surface more closely than the neighbouring simulations in the interaction sweep.

This shows that tetrahedral and octahedral conditions are two exact limits of the present $J_1$-$J_2$ model. Away from these limits, the relative penalties applied to first- and second-neighbour pairs vary continuously. The diamond- and rounded-square-shaped patterns can therefore be interpreted as intermediate regions of the same continuous pair-interaction landscape.

The cubic condition is not fully accounted for by $J_1$ and $J_2$ because the cube morphology requires constraints from higher-order pair interactions. A cube contains twelve second-neighbour edge pairs, twelve fourth-neighbour face-diagonal pairs and four sixth-neighbour body-diagonal pairs. Because the cubes overlap, each second-neighbour pair is shared by four cubes, each fourth-neighbour pair by two cubes and each sixth-neighbour pair by one cube. An interaction that is exactly equivalent to penalizing deviations from a balanced $Li_4M_4$ cube would therefore require interactions in the ratio $J_2$:$J_4$:$J_6$ = 4:2:1. Therefore, the $J_2$ only endpoint at $\lambda = 1$ consequently does not reproduce the complete cubic De Ridder surface in the same exact sense as the tetrahedral and octahedral limits. Nevertheless, $J_2$ acts along the twelve edges of each cube. Favouring unlike cations along these edges is sufficient to enrich balanced $Li_4M_4$ cubes and generate a cube-like diffuse-scattering morphology.

Finally, the De Ridder surfaces are three-dimensional surfaces which define where diffuse scattering is allowed to be present. At finite temperature, the local constraints are only partially satisfied, so the simulated diffuse intensity is broadened around the ideal surface.

**Supplementary Discussion 2:**

Following Sauvage and Parthé[1], we quantify the width of a polyhedral composition distribution relative to random cation mixing. For a polyhedron *P* containing $N_P$ cation sites, we denote this normalized variance as

$$R_P = \frac{\mathrm{Var}(n_\mathrm{M})}{N_P x_\mathrm{M}(1 - x_\mathrm{M})}$$

where $n_M$ is the number of M cations in the polyhedron and $x_M$ is the bulk M fraction. The denominator is the variance expected for a random binomial distribution. Consequently, $R_P$ = 1 indicates the same distribution width as random mixing, $R_P$ < 1 indicates a narrower distribution and $R_P$ > 1 indicates a broader distribution. The general cluster relationship derived by Sauvage and Parthé is equivalent to expressing $R_P$ in terms of the Warren–Cowley parameters for the pairs within each polyhedron. Using neighbour shells indexed by increasing distance on the complete fcc cation lattice gives

$$R_{\mathrm{tet}} = 1 + 3\alpha_1,$$
$$R_{\mathrm{oct}} = 1 + 4\alpha_1 + \alpha_2,$$
$$R_{\mathrm{cube}} = 1 + 3\alpha_2 + 3\alpha_4 + \alpha_6.$$

Here, $\alpha_r$ = 0 represents random mixing at neighbour shell *r*, $\alpha_r$ < 0 an unlike-cation preference and $\alpha_r$ > 0 a like-cation preference.
A tetrahedron contains only first-neighbour pairs. Positive $J_1$ therefore produces $\alpha_1$ < 0 and $R_{\mathrm{tet}}$ < 1, accounting for the strong enrichment of $Li_2M_2$ tetrahedra at $\lambda$ = 0. As $J_1$ is replaced by $J_2$, $\alpha_1$ approaches zero and the tetrahedral distribution approaches the random reference. Positive $J_1$ also suppresses the $Li_4$ tetrahedra that provide 0-TM migration environments. However, $R_{\mathrm{tet}}$ describes the overall distribution width and does not uniquely determine the $Li_4$ population.

An octahedron contains both first- and second-neighbour pairs. Either $\alpha_1$ < 0 or $\alpha_2$ < 0 can therefore reduce $R_{\mathrm{oct}}$ and enrich $Li_3M_3$ octahedra. Consistent with this, $R_{\mathrm{oct}}$ remains below unity throughout the complete interaction sweep and shows no unique minimum at the interaction ratio associated with the ideal octahedral De Ridder surface. A balanced octahedral distribution is consequently not an exclusive signature of octahedral-type scattering.

The cube contains second-neighbour edge pairs, fourth-neighbour face-diagonal pairs and sixth-neighbour body-diagonal pairs. Sauvage and Parthé expressed the corresponding relationship as $1 + 3\alpha_1 + 3\alpha_2 + \alpha_3 = 0$ for compositionally invariant cubes on a simple-cubic CsCl corner lattice. The same three distances correspond to the second-, fourth- and sixth-neighbour shells of the complete fcc cation lattice, giving $1 + 3\alpha_2 + 3\alpha_4 + \alpha_6 = 0$. This is the $R_{\mathrm{cube}}$ = 0 limit, in which every cube has the bulk cation composition.

Positive $J_2$ acts along the cube edges and favours edge-alternating $Li_4M_4$ arrangements. It can therefore produce balanced cubes while indirectly generating the accompanying fourth- and sixth-neighbour correlations. Explicit $J_4$ and $J_6$ interactions are not required for $R_{\mathrm{cube}}$ to approach zero. This explains why the cubic distribution becomes strongly enriched in $Li_4M_4$ environments in the $J_2$ -dominated simulations.

Conversely, the $J_1$-only structure has $R_{\mathrm{cube}}$ > 1, despite the cube containing no first-neighbour pairs. If $J_1$ had no influence on the cubic environments, $R_{\mathrm{cube}}$ would remain close to the random value as $J_2$ approaches zero, as indicated by the dotted line in Supplementary Fig. 3. Instead, the calculated $R_{\mathrm{cube}}$ increases to approximately 2.2 at $\lambda$ = 0. Satisfying $J_1$ across the network of overlapping local environments therefore indirectly correlates sites separated by the second-, fourth- and sixth-neighbour distances, producing the observed enrichment of $Li_8$ and $M_8$ cubes. This indirect correlation can be understood by considering an M cation and two of its first-neighbour sites, A and B, which are second neighbours of one another. Positive $J_1$ favours Li occupation of both A and B because each then forms an unlike M–Li first-neighbour pair with the central M cation. This simultaneously creates a Li–Li second-neighbour correlation between A and B. The equivalent Li-centred environment produces an M–M second-neighbour correlation. Thus, $J_1$ can generate like-cation correlations between cube vertices even though the cube contains no first-neighbour pairs, explaining the broadened cubic composition distribution in the $J_1$-only structure.

The normalized variance provides a compact measure of polyhedral compositional balance across the interaction sweep, but it is not a complete order parameter. Different histogram shapes and spatial arrangements can have the same $R_p$ , and $R_p$ does not determine the connectivity of neighbouring environments or the population of a particular composition such as $Li_4$. The complete polyhedral distributions, Warren–Cowley parameters and diffuse-scattering patterns therefore provide complementary descriptions of the simulated short-range order.

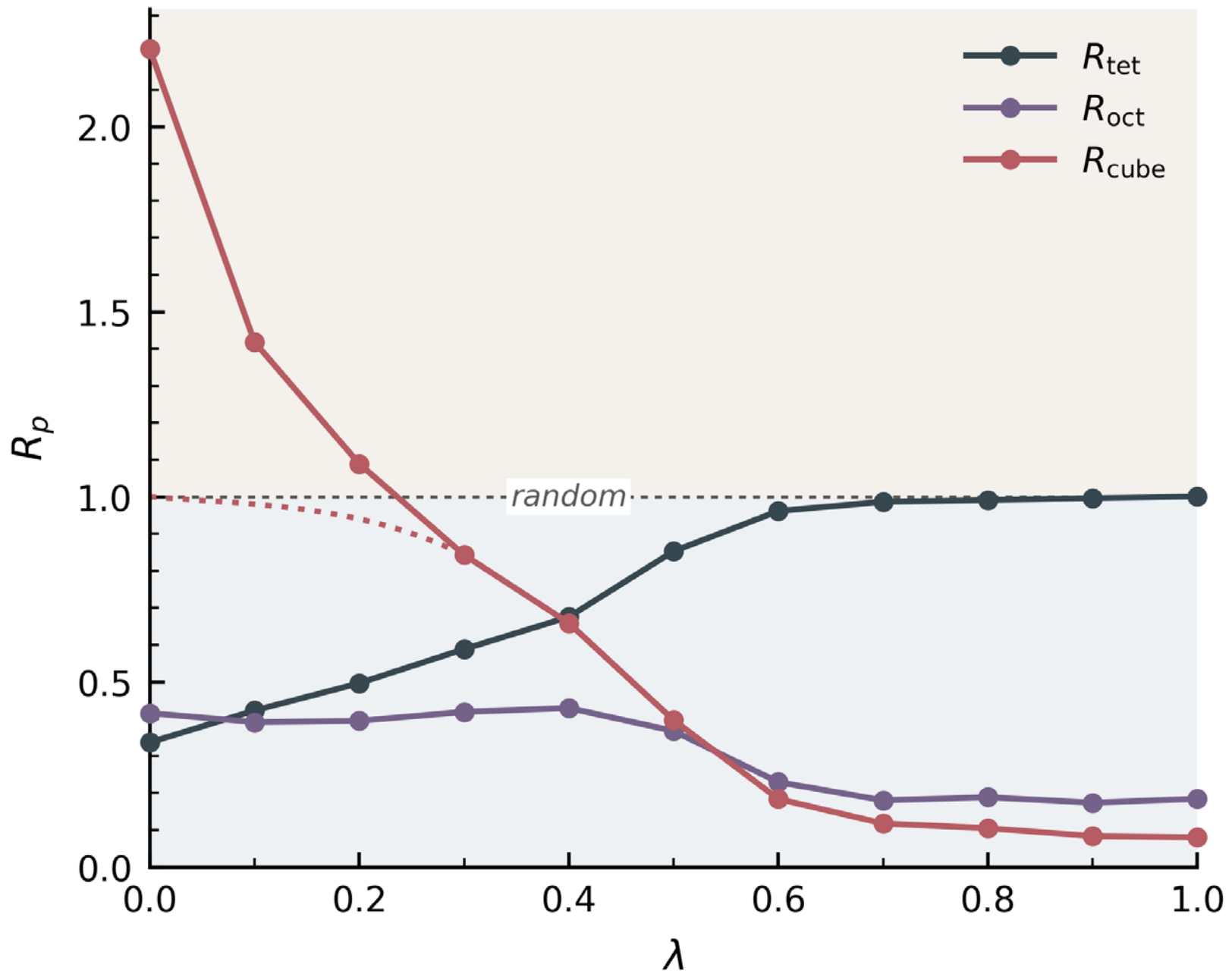

**Supplementary Fig. 3 | Dependence of local ordering on the balance of first- and second-neighbour interactions.** Warren–Cowley combinations $R_{\mathrm{tet}} = 1 + 3\alpha_1$, $R_{\mathrm{oct}} = 1 + 4\alpha_1 + \alpha_2$, and $R_{\mathrm{cube}} = 1 + 3\alpha_2 + 3\alpha_4 + \alpha_6$ are plotted against $\lambda = J_2 / (J_1 + J_2)$, at fixed total interaction strength $J_1 + J_2 = 2.32$ eV and $k_B T = 1$. Here, $\alpha_n$ denotes the $n^{\mathrm{th}}$-neighbour Warren–Cowley parameter. The horizontal dashed line marks the random reference, $R_p = 1$; shaded regions indicate clustering $R_p > 1$ and mixing $R_p < 1$. Symbols show calculated values, with solid lines joining adjacent points. The red dotted curve indicates the expected behaviour of $R_{\mathrm{cube}}$ in the absence of an indirect contribution from $J_1$.

**Supplementary Discussion 3:** Decoupling displacive and substitutional signatures in the 3D-ΔPDF

As shown by Roth & Iversen[1], the amplitude of a 3D-ΔPDF feature can be obtained by integrating the feature over a finite volume centred on the peak position:

$$\int_V \text{3D-}\Delta\text{PDF}\, d\boldsymbol{r} \propto \sum_{i,j|\boldsymbol{r}_{ij}\in V} \delta Z_i \delta Z_j \,.$$

The summation runs over all atomic pairs $(i, j)$ whose interatomic vectors $\boldsymbol{r}_{ij}$ fall within the integration volume centred on the 3D-ΔPDF feature of interest. Here, $\delta Z_i$ denotes the difference between the number of electrons associated with site $i$ in the real structure and that in the average structure.

The sign of the integrated peak amplitude directly reflects the nature of the substitutional correlation between the two sites. A positive amplitude indicates that sites $i$ and $j$ tend to be occupied simultaneously by elements that are either both heavier or both lighter than those present in the average structure. In contrast, a negative amplitude indicates a negative correlation, where occupation of one site by a heavier-than-average element is associated with occupation of the other site by a lighter-than-average element.

Thus, the sign of the integrated peak amplitude provides a direct measure of the sign of the underlying substitutional correlation, decoupled from any contributions arising from atomic displacements.

The integration boxes used for the data presented here were selected to be 7×7×7 voxels. Generally, the size of the integration boxes should be chosen such that they are large enough to enclose the feature of interest while being small enough as to not include excessive amounts of noise.[2] The integration box size can be evaluated by inspecting Supplementary Fig. 4. The resulting box integrated 3D-ΔPDF is shown in the Main Text.

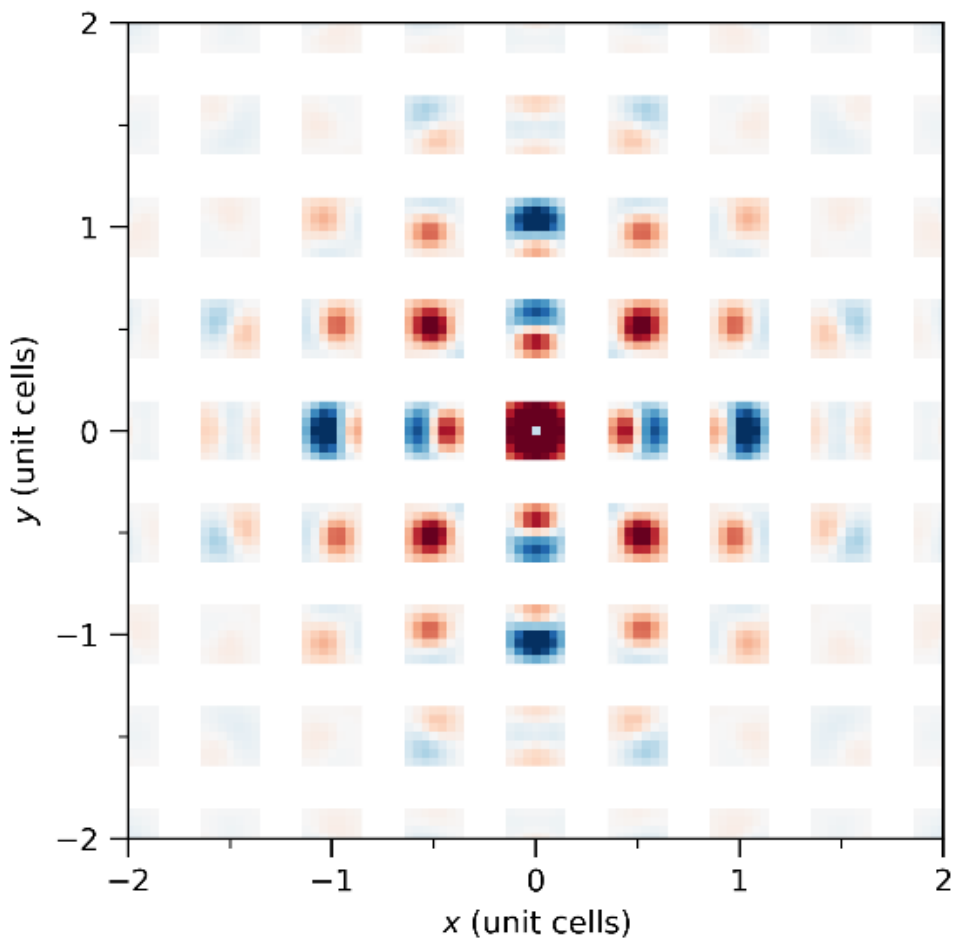

**Supplementary Fig. 4 | Evaluation of box size used for 3D-ΔPDF integration.** The figure shows the *xy*-plane of the 3D-ΔPDF with a mask overlaid to show the size of the integration boxes compared to the 3D-ΔPDF features.

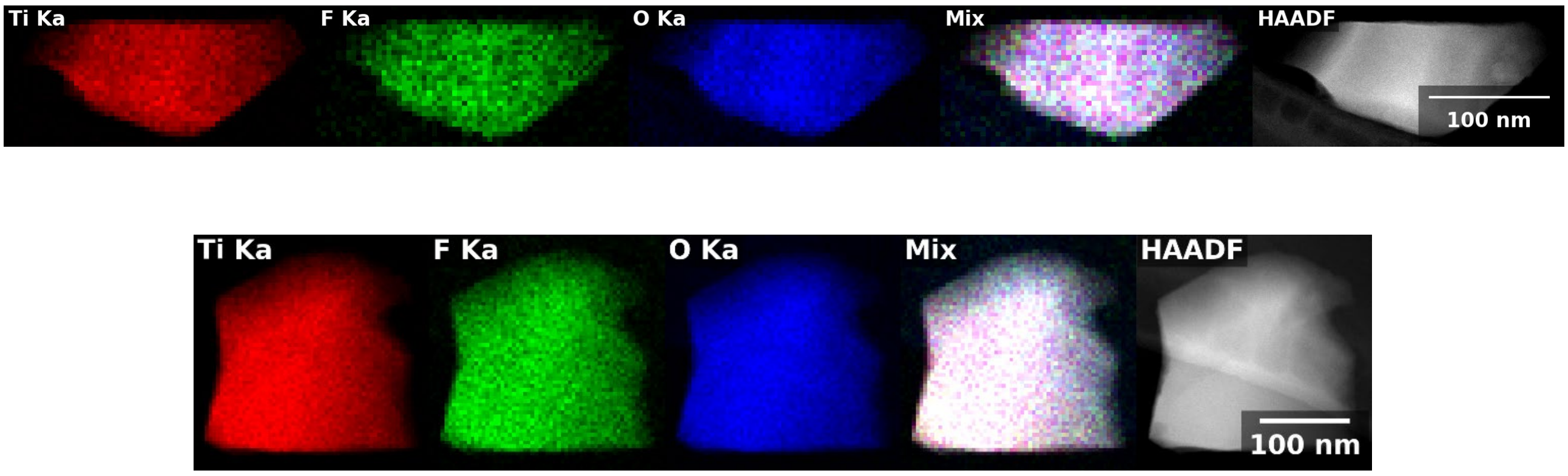

**Supplementary Fig. 5 | Elemental mapping of $Li_{2.5}TiO_3F_{0.5}$ particles.** Energy-dispersive X-ray maps showing the distributions of Ti (red), F (green) and O (blue), their overlays (Mix), and corresponding high-angle annular dark-field (HAADF) images. Top shows the crystallite used for the majority of the analyses. Bottom shows the larger crystal used in the comparison of Supplementary Fig. 10.

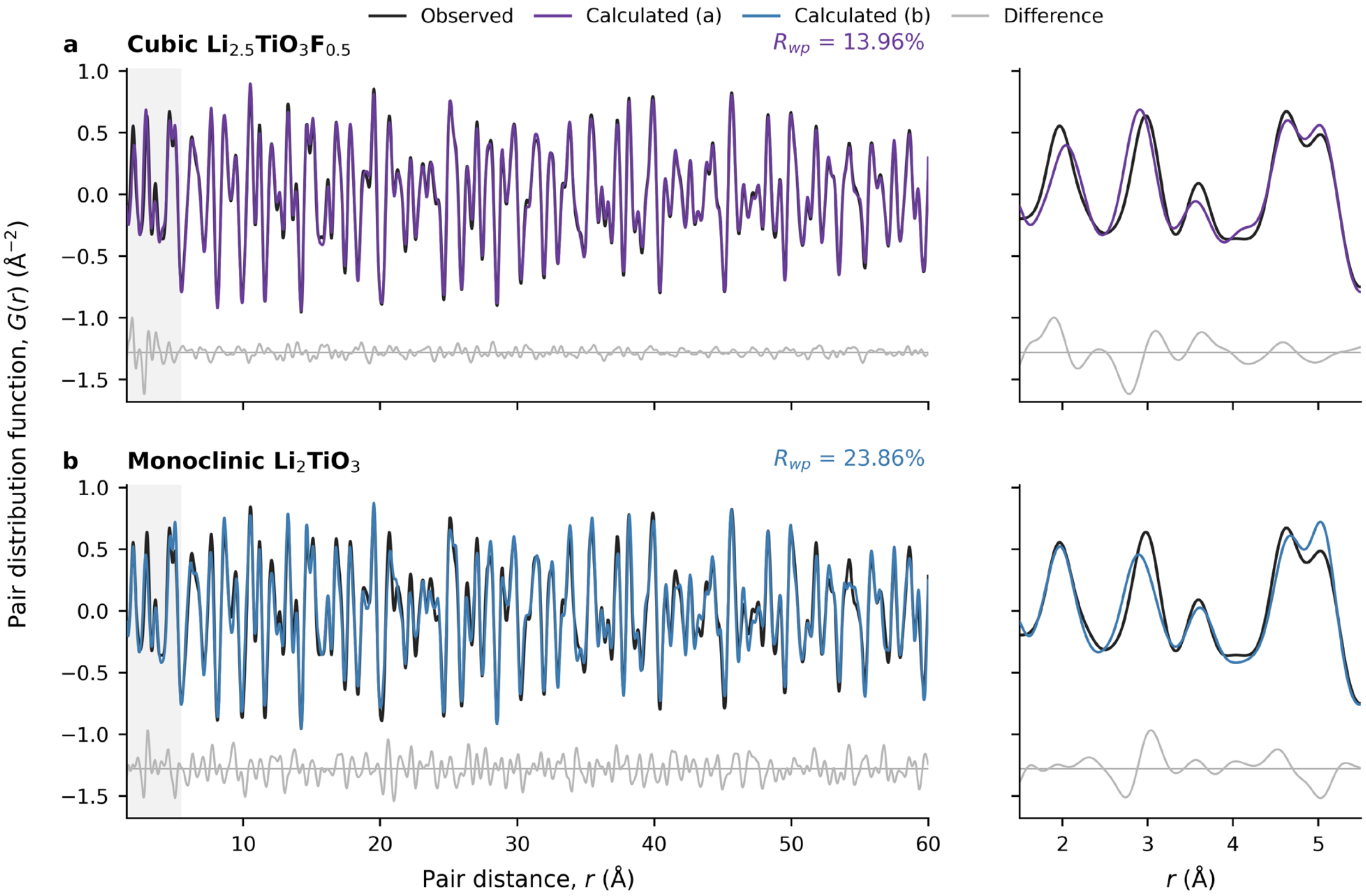

**Supplementary Fig. 6 | X-ray PDF refinements using $Li_{2.5}TiO_3F_{0.5}$ and impurity phase $Li_2TiO_3$.**

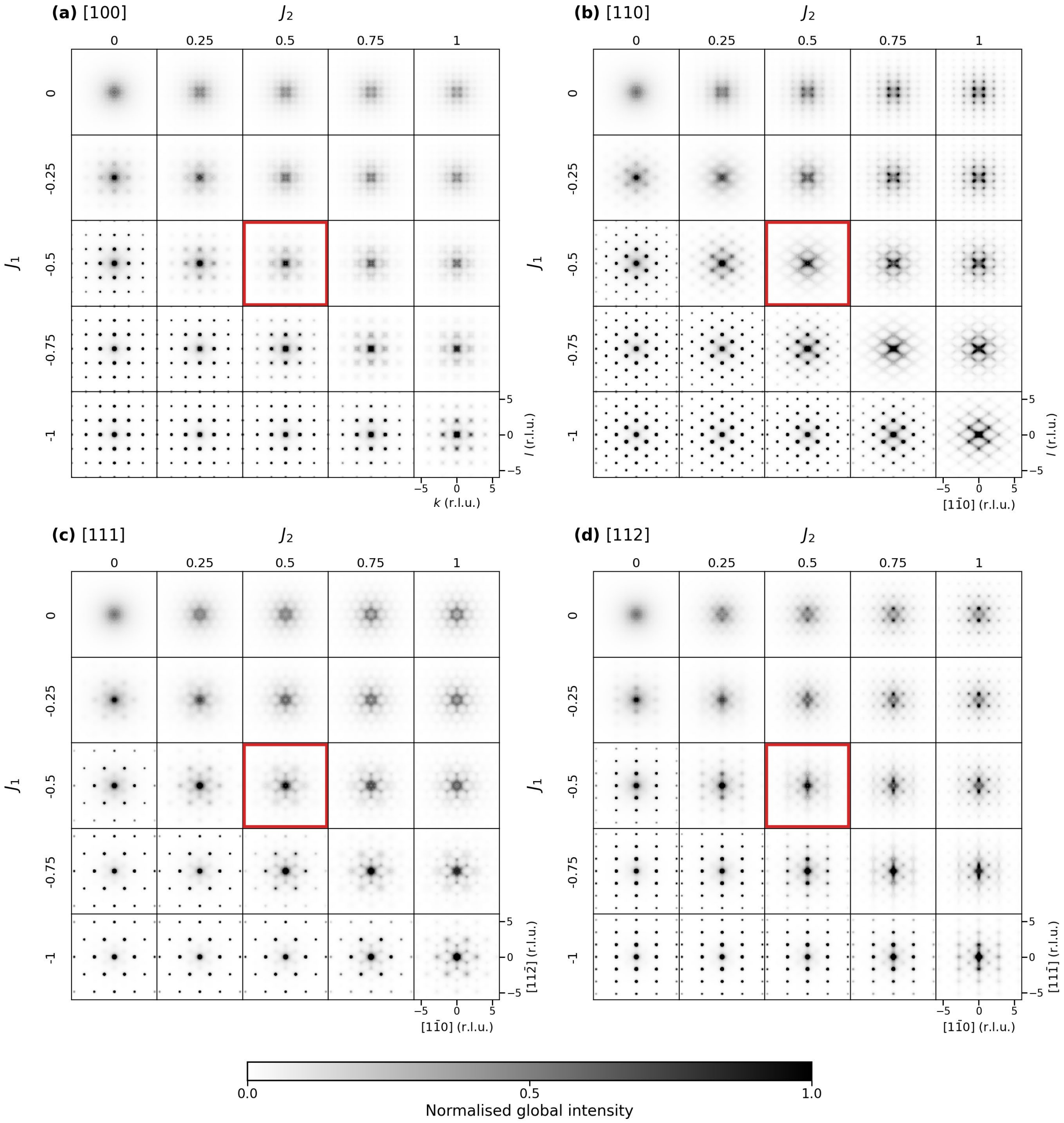

**Supplementary Fig. 7 | Sensitivity of the occupational diffuse scattering to the reduced interaction parameters.** Calculated reciprocal-space sections normal to (a) [100], (b) [110], (c) [111] and (d) [112] for a grid of reduced nearest- and next-nearest-neighbour parameters. All calculations employed the same composition, supercell, Monte Carlo protocol and intensity processing, and include occupational disorder only. The outlined cell denotes the representative parameter combination $J_1$, $J_2$ = -0.5, +0.5 used for subsequent displacement modelling. This was chosen by considering comparisons to the reconstructed experimental data shown in Fig. 1d. Neighbouring parameter combinations produce similar features, demonstrating that the selected values are non-unique phenomenological parameters rather than refined physical interaction energies. Intensities are normalized across the whole figure.

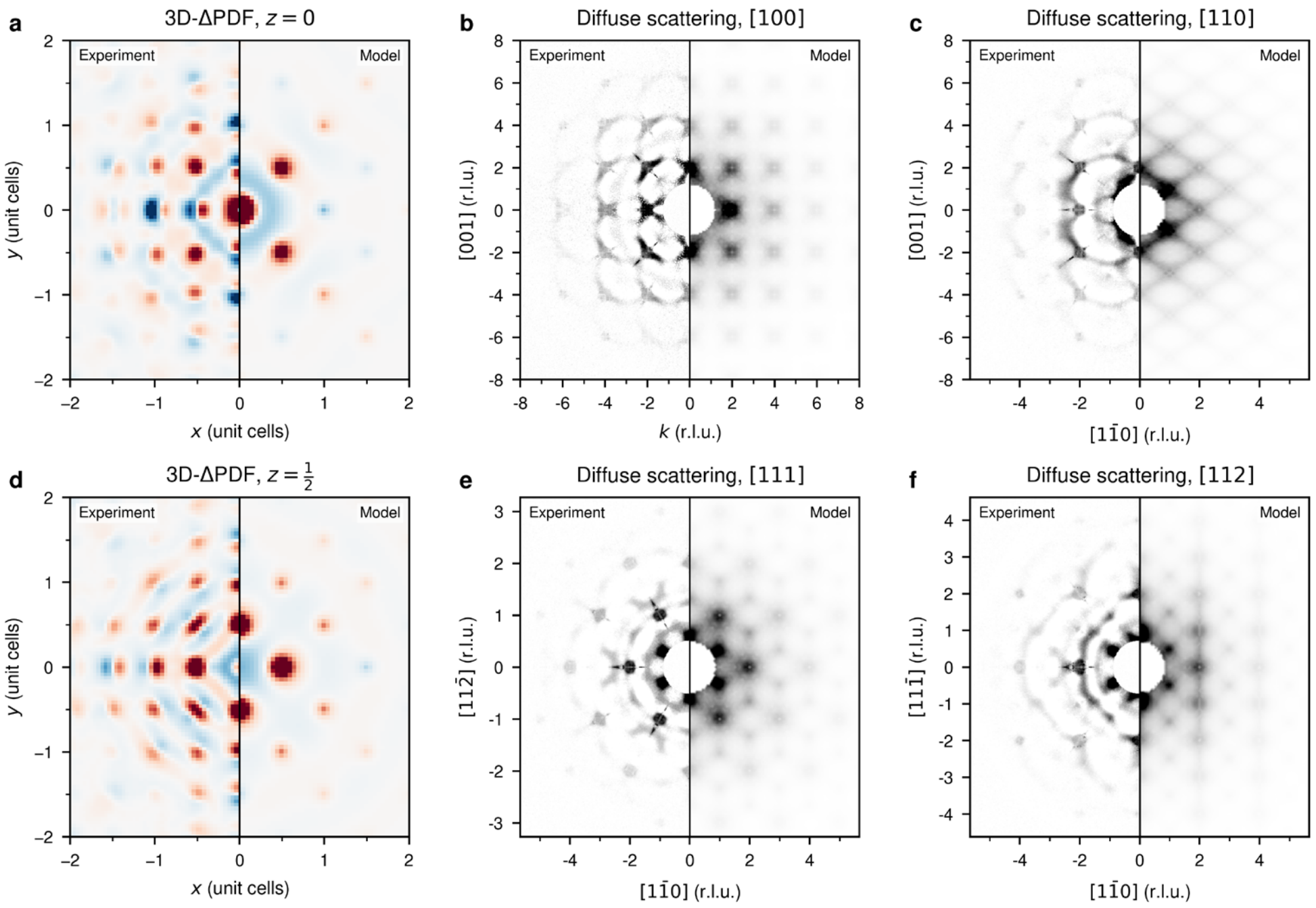

**Supplementary Fig. 8 | 3D-ΔPDF-informed structural model compared with experiment showing only the contribution from the occupational disorder.** Experimental data are shown on the left and model calculations on the right of each panel. a,d, 3D-ΔPDF sections at z = 0 (a) and z = ½ (d), with positive and negative intensity shown in red and blue, respectively. b,c,e,f, Diffuse-scattering sections normal to [100] (b), [110] (c), [111] (e) and [112] (f), with darker contrast indicating higher intensity. The model combines enhanced like-cation first-neighbour and unlike-cation second-neighbour populations. Real-space coordinates are expressed in conventional cubic unit cells and reciprocal-space coordinates in reciprocal-lattice units

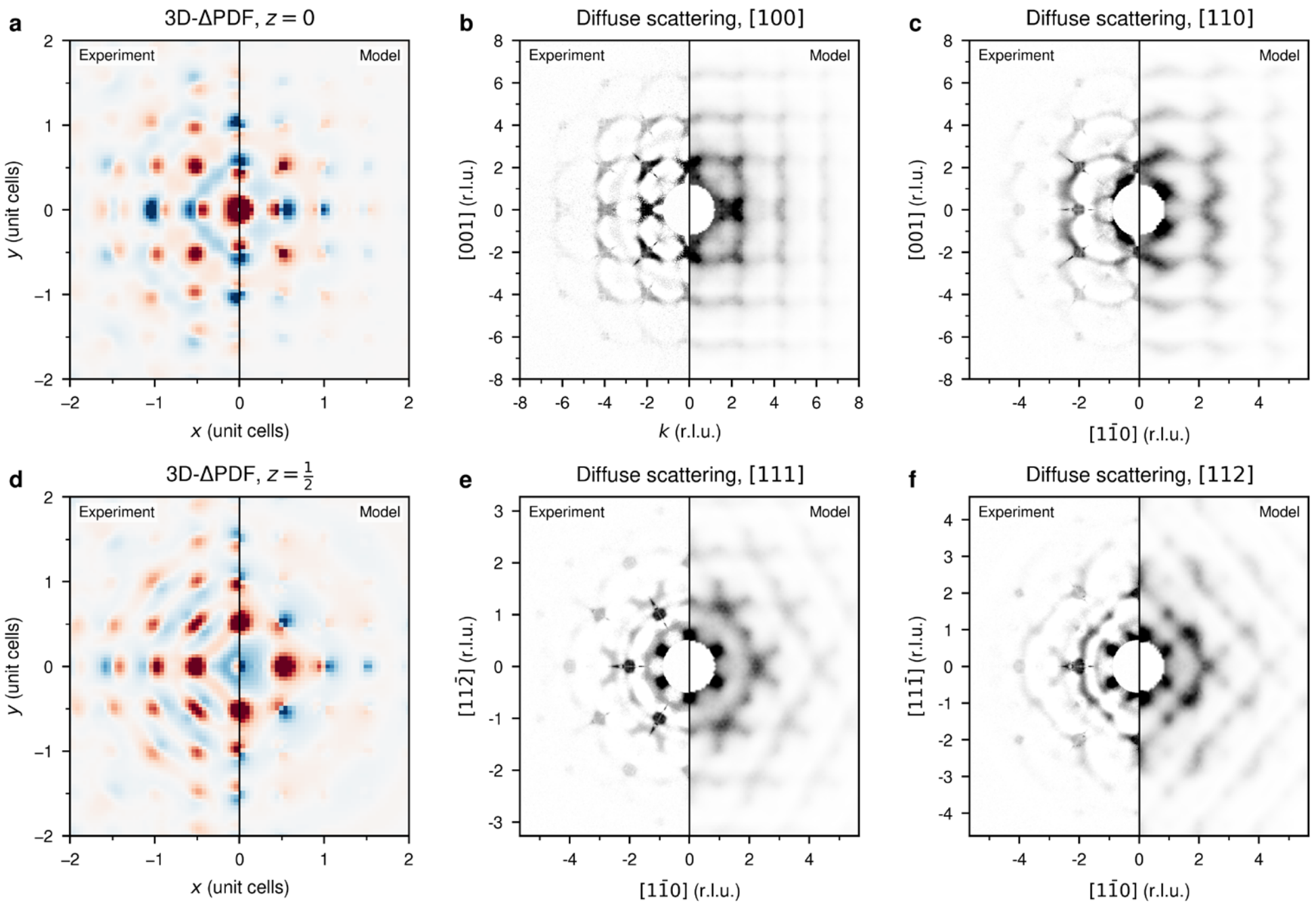

**Supplementary Fig. 9 | 3D-ΔPDF-informed structural model with composition $Li_{2.43}TiO_3F_{0.43}$ compared with experiment.** Experimental data are shown on the left and model calculations on the right of each panel. a,d, 3D-ΔPDF sections at z = 0 (a) and z = ½ (d), with positive and negative intensity shown in red and blue, respectively. b,c,e,f, Diffuse-scattering sections normal to [100] (b), [110] (c), [111] (e) and [112] (f), with darker contrast indicating higher intensity. The model combines enhanced like-cation first-neighbour and unlike-cation second-neighbour populations correlated anion and cation displacements. Real-space coordinates are expressed in conventional cubic unit cells and reciprocal-space coordinates in reciprocal-lattice units. The composition used here is the one suggested by STEM-EDX measurements and shows little sensitivity in the diffuse scattering prediction from small deviations from the nominal $Li_{2.5}TiO_3F_{0.5}$.

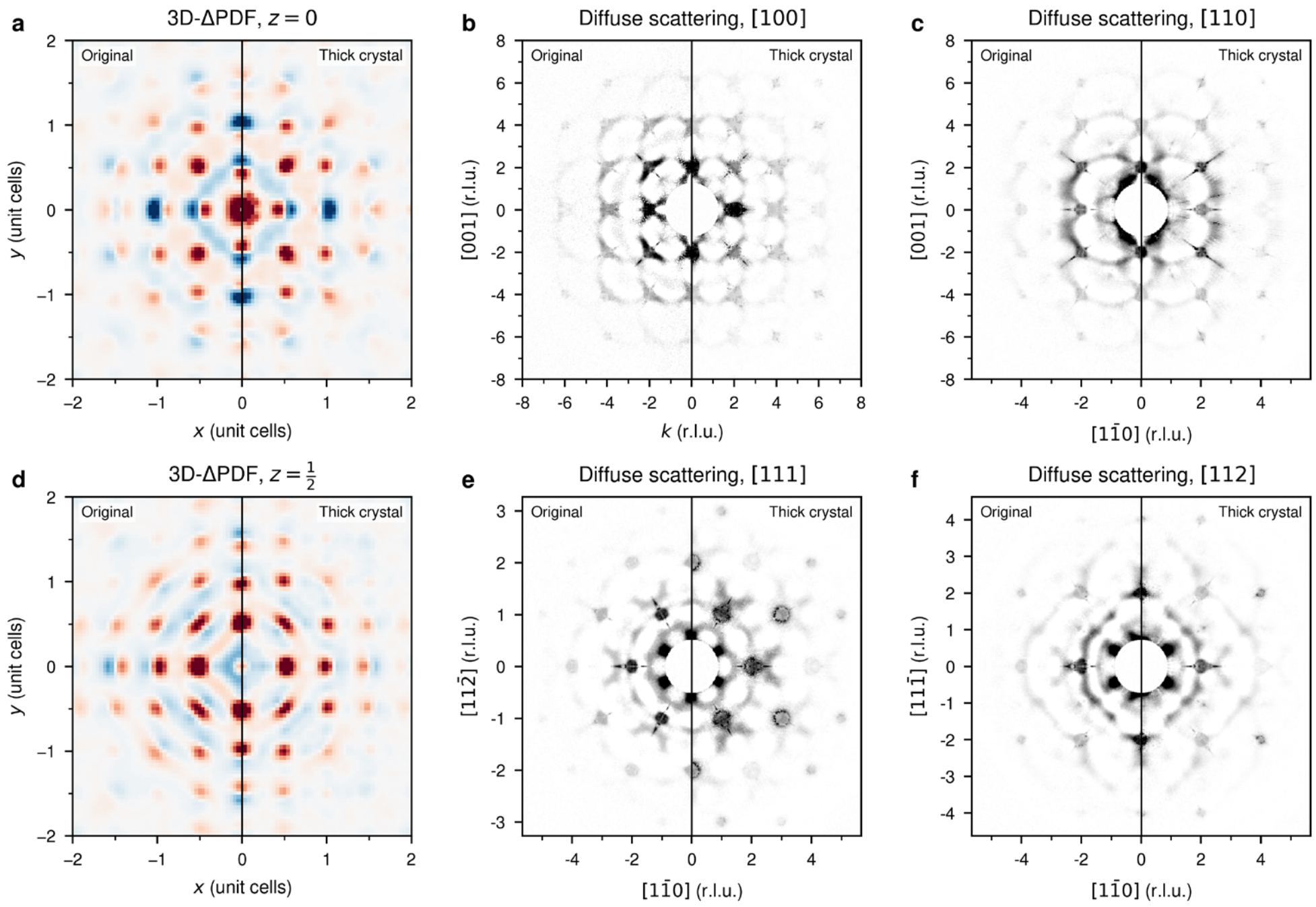

**Supplementary Fig. 10 | Comparison of two datasets collected on different crystallites.** The left of the figure shows the data used in the analysis of the main article and the right shows the data of the secondary crystal. a,d, 3D-ΔPDF sections at z = 0 (a) and z = ½ (d), with positive and negative intensity shown in red and blue, respectively. b,c,e,f, Diffuse-scattering sections normal to [100] (b), [110] (c), [111] (e) and [112] (f), with darker contrast indicating higher intensity.

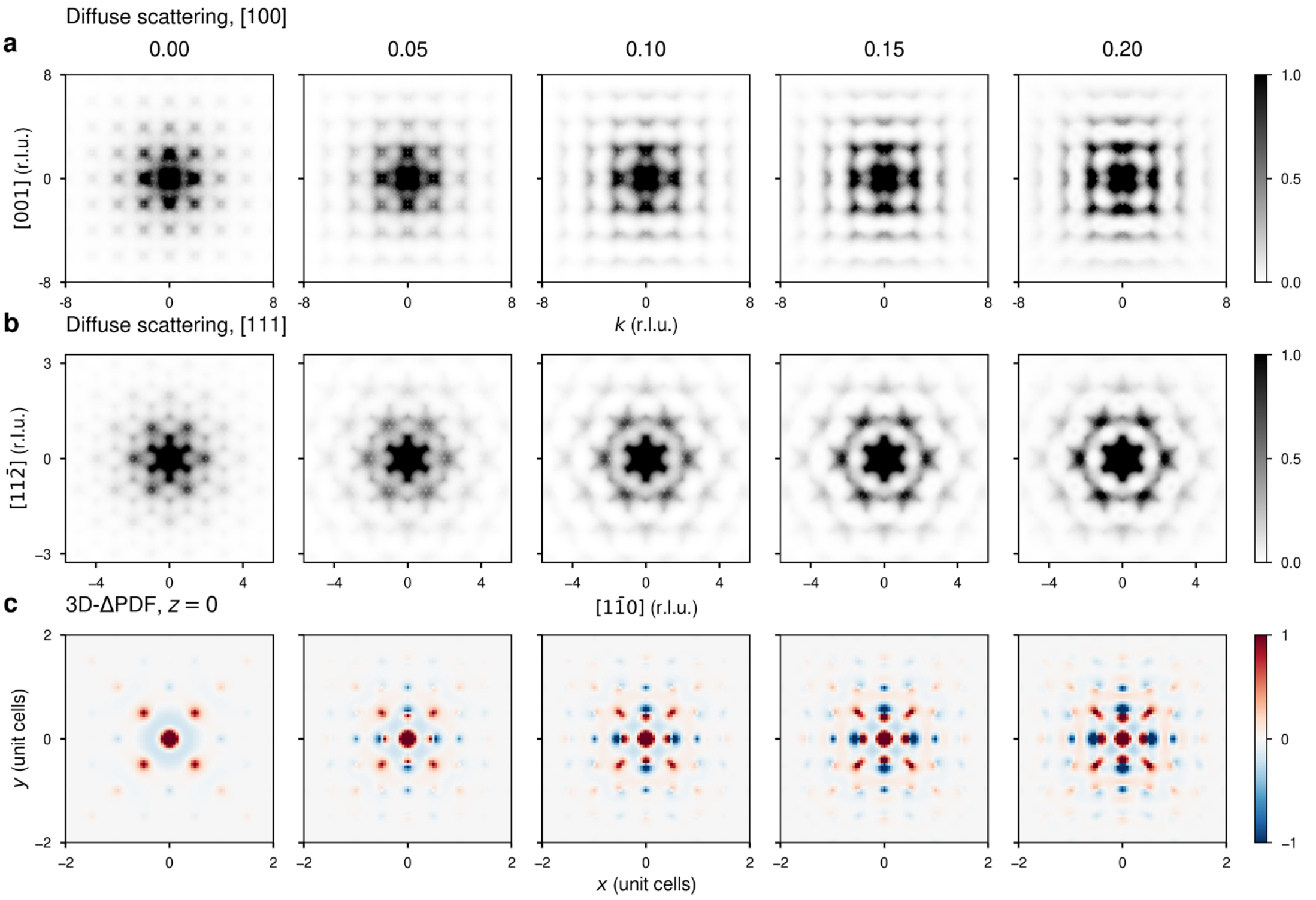

**Supplementary Fig. 11 | Ti–X displacement sweep.** Effect of increasing nearest-neighbour Ti–X contraction on calculated **a**, [100] diffuse scattering; **b**, [111] diffuse scattering; and **c**, the xy-plane 3D-ΔPDF at ($z$ = 0). Columns show displacement amplitudes of 0, 0.05, 0.10, 0.15 and 0.20 Å. Anions move towards neighbouring Ti sites, with the other displacement modes inactive.

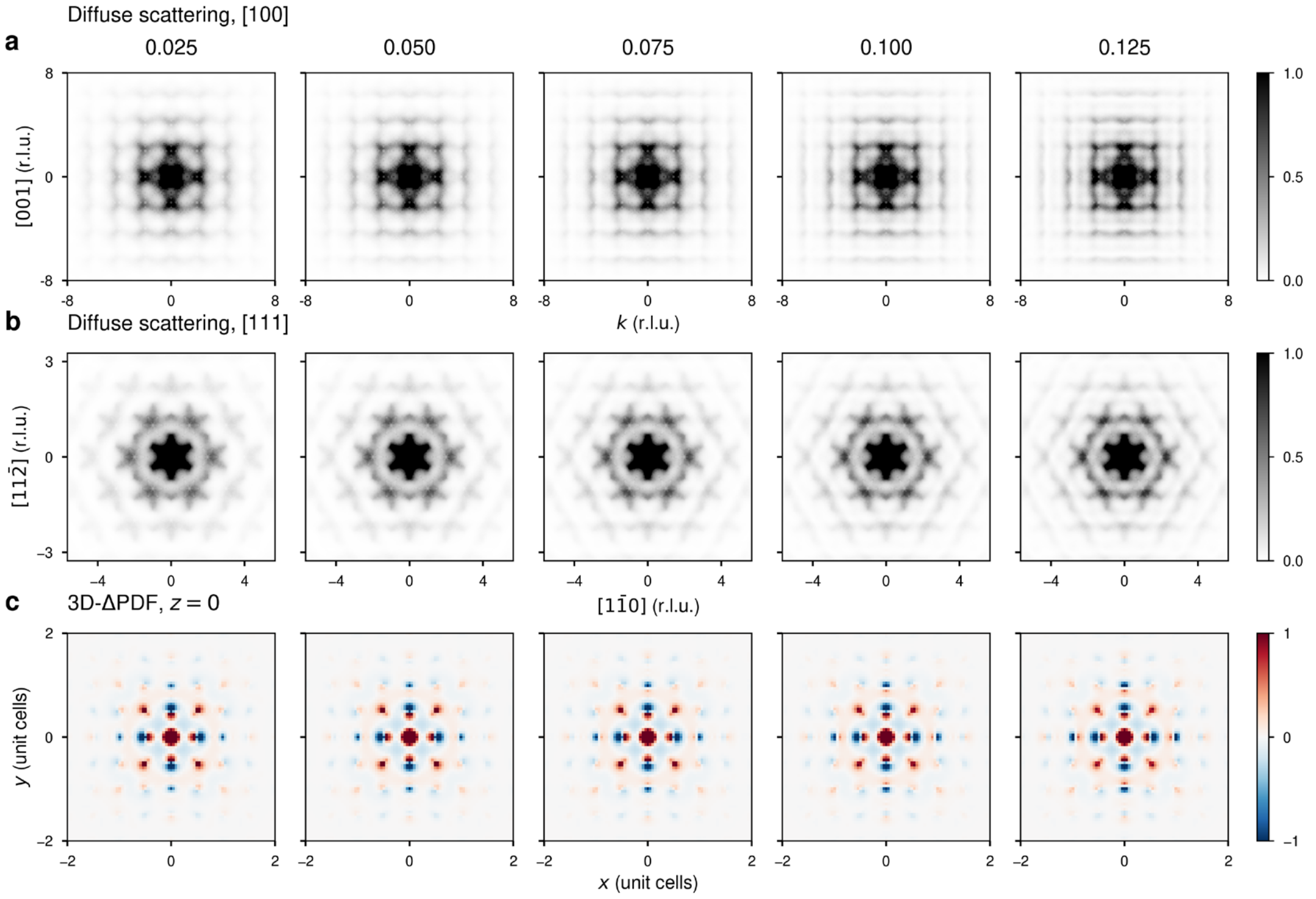

**Supplementary Fig. 12 | Ti–Li second-neighbour displacement sweep.** Calculated **a**, [100] diffuse scattering; **b**, [111] diffuse scattering; and **c**, the xy-plane 3D-ΔPDF at ($z$ = 0), for increasing Ti–Li second-neighbour separation along <100>. Columns show displacement amplitudes of 0.025, 0.050, 0.075, 0.100 and 0.125 Å per pair contribution. Both cations move, while Ti–X contraction and Ti–Ti nearest-neighbour separation remain fixed at 0.100 and 0.025 Å, respectively.

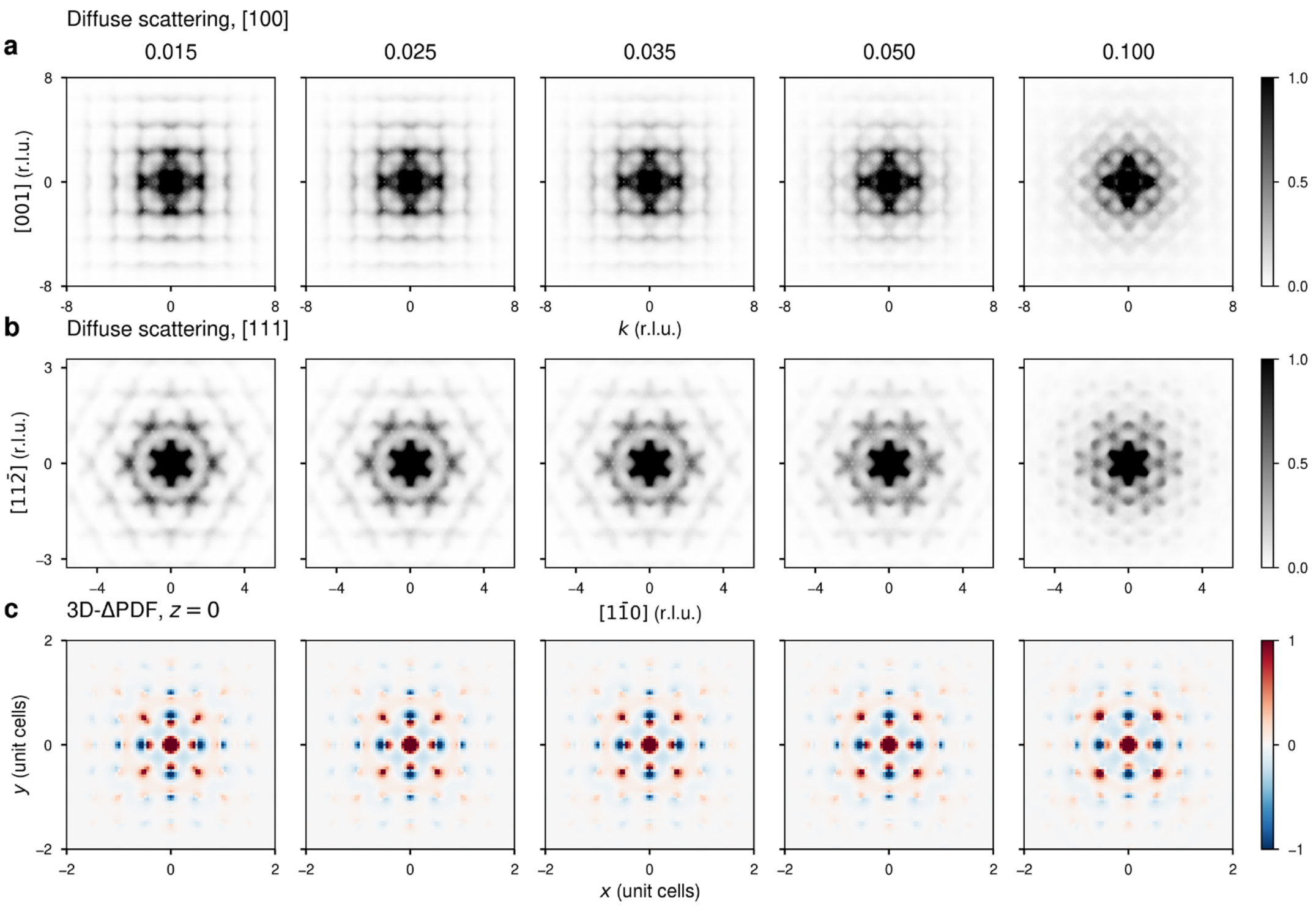

**Supplementary Fig. 13 | Ti–Ti nearest-neighbour displacement sweep.** Calculated **a**, [100] diffuse scattering; **b**, [111] diffuse scattering; and **c**, the xy-plane 3D-ΔPDF at ($z$ = 0), for increasing Ti–Ti nearest-neighbour separation along <110>. Columns show displacement amplitudes of 0.015, 0.025, 0.035, 0.050 and 0.100 Å per pair contribution. Ti–X contraction and Ti–Li second-neighbour separation remain fixed at 0.100 and 0.075 Å, respectively.

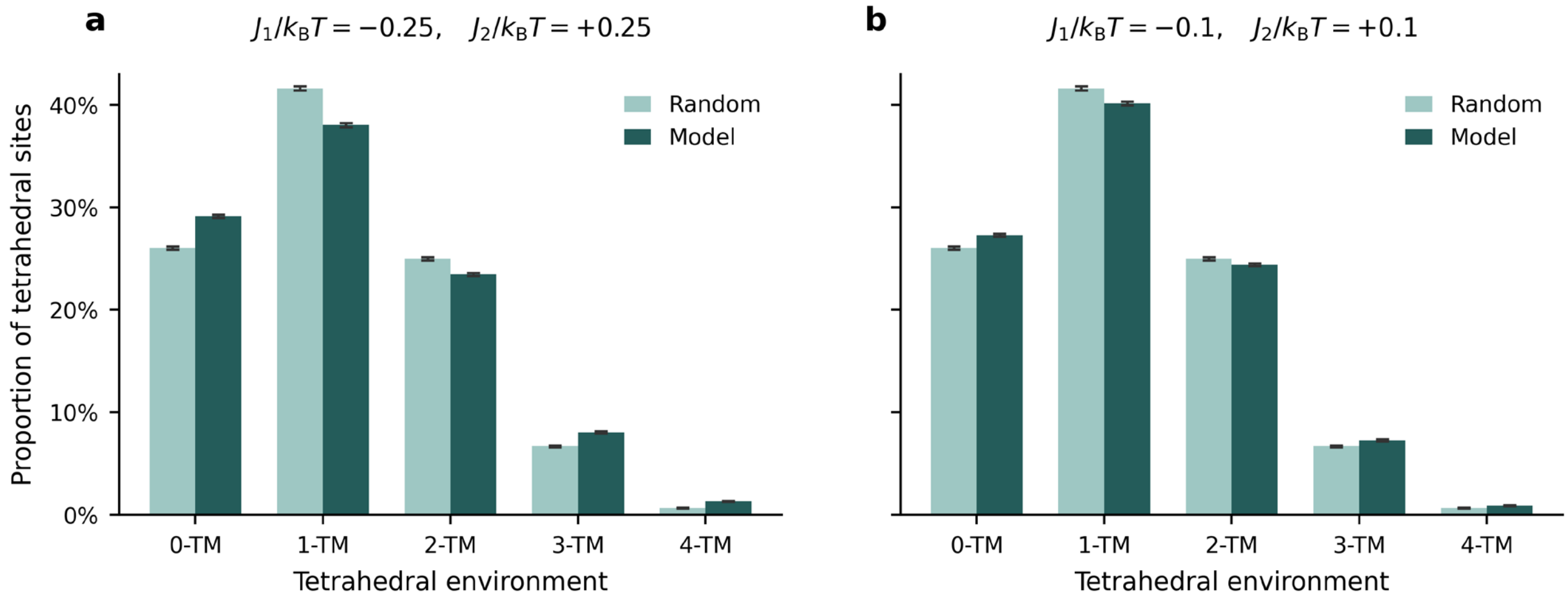

**Supplementary Fig. 14 | Population of 0-TM tetrahedral sites for weaker *J*.** Proportions of *n*-TM tetrahedral environments in the pair-interaction model compared with random cation mixing at the same composition. The occupations were determined from a Monte Carlo simulation using the pair interaction model discussed in the main text using $J_1$, $J_2$ = –0.25, +0.25 and –0.1, +0.1.